\documentclass[2column]{aa}  

\usepackage{epstopdf}
\usepackage{graphicx}
\usepackage{txfonts}
\usepackage{natbib}
\usepackage{multirow}
\usepackage{pdflscape}
\usepackage{lscape}

\begin{document} 

\title{Redshifted X-rays from the material accreting onto TW~Hya: \\
Evidence of a low-latitude accretion spot}

\titlerunning{Redshifted X-rays from TW~Hya: A low-latitude accretion spot}

\author{C. Argiroffi\inst{1,2}, J.~J. Drake\inst{3}, R. Bonito\inst{2}, S. Orlando\inst{2}, G. Peres\inst{1,2}, M. Miceli\inst{1,2}}
\authorrunning{C. Argiroffi et al.}

\institute{
Dip. di Fisica e Chimica, Universit\`a di Palermo, Piazza del Parlamento 1, 90134, Palermo, Italy, \email{argi@astropa.unipa.it}
\and
INAF - Osservatorio Astronomico di Palermo, Piazza del Parlamento 1, 90134, Palermo, Italy
\and
Smithsonian Astrophysical Observatory, MS-3, 60 Garden Street, Cambridge, MA 02138, USA} 
\date{Received 9 June 2017; accepted 9 August 2017}

\abstract
{High resolution spectroscopy, providing constraints on plasma motions and temperatures, is a powerful means to investigate the structure of accretion streams in classical T~Tauri stars (CTTS). In particular, the accretion shock region, where the accreting material is heated to temperatures of a few million degrees as it continues its inward bulk motion, can be probed by X-ray spectroscopy.}
{In an attempt to detect for the first time the motion of this X-ray-emitting post-shock material, we searched for a Doppler shift in the deep \textit{Chandra} High Energy Transmission Grating observation of the CTTS TW~Hya. This test should unveil the nature of this X-ray emitting plasma component in CTTS and constrain the accretion stream geometry.}
{We searched for a Doppler shift in the X-ray emission from TW~Hya with two different methods: by measuring the position of a selected sample of emission lines and by fitting the whole TW~Hya X-ray spectrum, allowing the line-of-sight velocity to vary.}
{We found that the plasma at $T\sim 2-4$\,MK has a line-of-sight velocity of $38.3\pm5.1\,{\rm km\,s^{-1}}$ with respect to the stellar photosphere. This result definitively confirms that this X-ray-emitting material originates in the post-shock region, at the base of the accretion stream, and not in coronal structures. The comparison of the observed velocity along the line of sight, $38.3\pm5.1\,{\rm km\,s^{-1}}$, with the inferred intrinsic velocity of the post shock of TW~Hya, $v_{post}\approx 110-120\,{\rm km\,s^{-1}}$, indicates that the footpoints of the accretion streams on TW~Hya are located at low latitudes on the stellar surface.}
{Our results indicate that complex magnetic field geometries, such as those of TW~Hya, permit low-latitude accretion spots. Moreover, since on TW~Hya the redshift of the soft X-ray emission is very similar to that of the narrow component of the \ion{C}{iv} resonance doublet at 1550\,\AA,\, then the plasma at $2-4$\,MK and that at 0.1\,MK likely originate in the same post-shock regions.}


\maketitle

\section{Introduction}

Low-mass stars ($M\lesssim2\,M_{\odot}$), at the initial stages of their pre-main-sequence evolution, are surrounded by circumstellar disks from which they accrete material. During this phase they are classified as classical T~Tauri stars (CTTS). The accretion phase persists for a few Myr, during which the circumstellar disk gradually exhausts its gas and dust content through accretion itself, photoevaporation, and planet formation \citep{Armitage2011,WilliamsCieza2011}. During the accretion phase, the star exchanges mass, angular momentum, and energy with its surrounding environment. These exchanges are thought to affect the stellar evolution, in terms of rotation, internal structure, bolometric luminosity, and lithium abundance, well beyond the accretion phase itself \citep{BouvierForestini1997,BaraffeChabrier2009,BaraffeChabrier2010}. 

Both observational and theoretical evidence clearly indicates that the magnetic field plays a major role in the accretion process of CTTS \citep[see][and references therein]{RomanovaOwocki2015,HartmannHerczeg2016}. The so-called magnetospheric accretion model predicts that the stellar magnetic field disrupts the inner disk, loads disk material in its flux tubes, and guides this material in its fall toward the star \citep[e.g.,][]{Koenigl1991,CalvetHartmann1992,HartmannHewett1994}. Because of the typical gravitational wells of CTTS, the accreting material accelerates up to $300-500\,{\rm km\,s^{-1}}$ and terminates its fall impacting onto the denser stellar atmosphere. There, at the footpoints of these accretion streams, strong shocks convert the bulk of the kinetic energy of the accreting material into thermal energy \citep{CalvetGullbring1998} and generate hot spots on the stellar surface. Because of the high pre-shock velocities, the post-shock material, located just behind the shock surface, is expected to have temperatures of a few MK, and hence to radiate mainly in the X-ray band \citep{Ulrich1976,Koenigl1991,Gullbring1994}. The observed emission of these accretion spots peaks instead in the UV and optical bands, indicating temperatures of $\sim10^4$\,K \citep{InglebyCalvet2013}. These temperatures, cooler than that expected, are thought to be related to a significant reprocessing of the primary X-rays into longer wavelength photons in the accretion-shock region itself. In fact, local X-ray absorption due to the surrounding stellar atmosphere and pre-shock material, and their consequent heating, could be substantial \citep{Drake2005,SaccoOrlando2010,BonitoOrlando2014,CostaOrlando2017}.

However, despite this expected local absorption, recent findings hint that some of the primary X-ray photons are in any case able to escape from the accretion-shock region. High resolution X-ray spectroscopy ($R\sim1000$) of CTTS in fact proved the existence of a cool ($T\sim3$\,MK) and high-density ($n_{\rm e}\sim10^{12}\,{\rm cm^{-3}}$) plasma component \citep[e.g.,][]{KastnerHuenemoerder2002,StelzerSchmitt2004,SchmittRobrade2005,GuntherLiefke2006}, as well as high-temperature and low-density coronal plasmas typical for low-mass stars. Several hydrodynamic (HD) and magnetohydrodynamic (MHD) models investigated the physical conditions of the accretion-shock region \citep[e.g.,][]{SaccoArgiroffi2008,SaccoOrlando2010,OrlandoSacco2010,OrlandoBonito2013,MatsakosChieze2013}, indicating that this high-density plasma at a few MK, which was never observed in non-accreting low-mass stars, is exactly what would be produced in an accretion shock on a CTTS. Recently, indications of X-rays emitted in accretion shocks were also obtained with low resolution X-ray spectroscopy ($R\sim40$), thanks to simultaneous optical monitoring \citep{GuarcelloFlaccomio2017}. All these findings make soft X-rays one of the most direct probes of the accretion-shock region. The diagnostic power offered by soft X-rays from CTTS is illustrated by their power to infer the densities and velocities of the stream \citep[e.g.,][]{KastnerHuenemoerder2002,StelzerSchmitt2004,GuntherLiefke2006,ArgiroffiMaggio2007}, the peculiar chemical composition of the inner disk material \citep{DrakeTesta2005}, indications on the accretion stream geometry \citep{ArgiroffiFlaccomio2011,ArgiroffiMaggio2012}, and the intrinsic variability of the accretion stream properties \citep{BrickhouseCranmer2012}.

Beside several theoretical and observational lines of evidence supporting the hypothesis that soft X-rays in CTTS come from plasma heated in the accretion shock, other results appear to conflict with this scenario: the lack of correlation between the soft X-ray excess with UV accretion indicators \citep{GuedelTelleschi2007}, the lack of a high-density plasma component on the CTTS T~Tauri \citep{GuedelSkinner2007}, and the density versus temperature pattern of X-ray emitting plasma \citep{BrickhouseCranmer2010}. Starting from these findings, alternative scenarios were proposed, suggesting that the cool plasma in CTTS could also be composed of coronal structures modified, cooled, or fed by accretion \citep{GuedelTelleschi2007,BrickhouseCranmer2010,SchneiderGunther2017}.

While the most important mechanisms governing magnetospheric accretion are understood, the geometry of the magnetosphere is still poorly constrained. Modeling of accretion onto CTTS usually assumes that disk disruption and material loading is due to the dipole component of the magnetic field \citep[e.g.,][]{HartmannHewett1994,RomanovaUstyugova2003,RomanovaLong2011} because this component is expected to be dominant at large distance. However, the geometry of the accretion streams, and consequently their footpoint locations, depends strongly on the magnetic configuration near the stellar surface. A dipole component that also dominates near the stellar surface would anchor the accretion stream footpoints at high latitudes \citep{RomanovaUstyugova2004}. Conversely, surface magnetograms indicate that CTTS have complex magnetic field configurations that are often characterized by intense higher order components \citep[e.g.,][]{GregoryMatt2008,DonatiLandstreet2009}. Higher order components could instead generate accretion streams terminating at low latitude on the stellar surface \citep{GregoryJardine2006}.

In addition to the magnetic field, accretion geometry could also be related to the inner structure of circumstellar disks. At ages of a few Myr, circumstellar disks often show complex and asymmetrical structures, such as inner disk warps and misalignments \citep[e.g.,][]{HashimotoTamura2011,ApaiSchneider2015,DebesPoteet2017}, which are possibly due to ongoing planet formation. These asymmetries, if they extend to the inner disk rim, could affect the accretion stream geometry.

To provide constraints on the accretion geometry, we test in this work one basic prediction of accretion-related X-rays that has not been investigated yet. The velocity jump predicted for strong-shock conditions across the shock surface, at the base of the accretion stream, is $v_{post}=v_{pre}/4$ \citep{ZeldovichRaizer1967}, where $v_{post}$ and $v_{pre}$ indicate the post- and pre-shock velocities, respectively. Therefore the plasma located just behind the shock surface is expected to have a downward bulk velocity $v_{post}$ of $\sim100\,{\rm km\,s^{-1}}$ within the stellar atmosphere\footnote{This expected value for $v_{post}$ is based on several observational and theoretical arguments; the most stringent argument is the fact that $v_{pre}$ of at least $300\,{\rm km\,s^{-1}}$ (corresponding to $v_{post}\approx75\,{\rm km\,s^{-1}}$) are needed to have X-ray emission from the post-shock plasma \citep{SaccoOrlando2010}.}. The emerging X-ray spectrum should therefore be redshifted with respect to the stellar photosphere. Hydrodynamic and MHD simulations have also confirmed this. In fact, even if the post-shock region is characterized by oscillations due to the classical radiative shock instability that is not detected \citep[e.g.,][]{DrakeRatzlaff2009}, the bulk post-shock velocity is still expected to vary between $\sim v_{pre}/8$ and $\sim v_{pre}/2$ \citep{SaccoArgiroffi2008,SaccoOrlando2010}. Performing temporal and spatial integration of the shock region, and considering the local absorption, the expected redshift should still be detectable with the present-day X-ray spectrometers, depending on the viewing angle under which the shock region is observed \citep[][Bonito et al., in prep.]{BonitoOrlando2014}.

Our aim is therefore to search for a Doppler shift in the X-ray emission from CTTS to measure the bulk velocity of the plasma components at a few MK with respect to the stellar photosphere. First of all, the detection of a redshift would be definitive evidence that the high-density plasma in CTTS is indeed located in the post-shock region and not in accretion-related coronal structures. The detected velocity could also provide constraints on the inclination angle under which the shock region is observed, and hence on the geometry of the accretion. Finally, the comparison of any detected velocity of the X-emitting plasma with that obtained from far UV lines also ascribed to the shock region \citep[e.g.,][]{ArdilaHerczeg2013}, also allows us to construct a coherent picture of the accretion-shock region in CTTS. To this aim, we selected the nearest CTTS, TW~Hya, because its almost pole-on orientation minimizes rotational modulation effects and maximizes the possibility to observe redshifts in case of high-latitude accretion shocks. Moreover, also because of its proximity and hence its high X-ray flux, TW~Hya was targeted for a Large Observing Project with the {\it Chandra} High Energy Transmission Gratings Spectrometer (HETGS), the instrument that currently provides the highest spectral resolution in the soft X-ray band.

We summarize the main properties of TW~Hya in Sect.~\ref{twhyaprop}. In Section ~\ref{chandraprop} we scrutinize the expected resolution of the {\it Chandra}/HETGS, and in Sect.~\ref{chandraprop} we detail the analyzed data sets. In Section~\ref{methods}, we present the two methods adopted to measure X-ray Doppler shifts, reporting the obtained results in Sect.~\ref{results}, and discussing these results in Sect.~\ref{discussion}.

\begin{figure*}
\centering
\includegraphics[width=17cm]{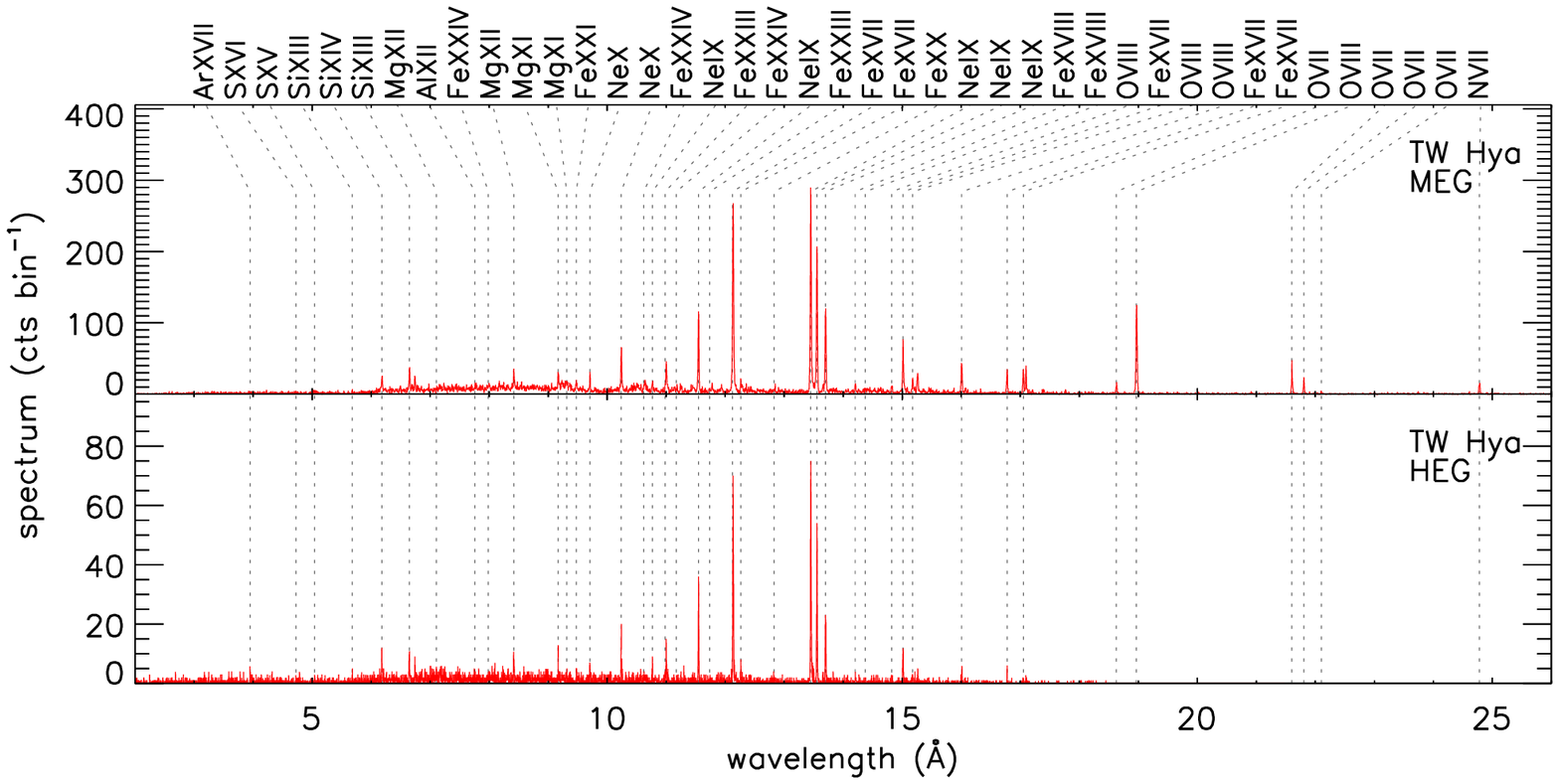}
\caption{MEG and HEG spectra of TW~Hya obtained during the 2007 observation (ObsID 7435+7436+7437).}
\label{plothetgspec}
\end{figure*}

\section{TW~Hya}
\label{twhyaprop}

Located at 59.5\,pc \citep{GaiaCollaboration2016}, TW~Hya is an 8 Myr old CTTS and the eponymous member of the TW~Hya Association \citep{KastnerZuckerman1997,WebbZuckerman1999,DonaldsonWeinberger2016}. The spectral type classification of TW\ Hya is uncertain, ranging from K7 to M2.5 depending on the wavelength band in which the classification is made \citep[see ][and references therein]{VaccaSandell2011,DebesJangCondell2013,HerczegHillenbrand2014}. As a consequence the mass and radius estimates of TW~Hya are also uncertain, with $M_{\star}$ ranging from $\sim 0.5$ to $\sim 0.8\,M_{\odot}$ and $R_{\star}$ ranging from $\sim 0.8$ to $\sim 1.1\,R_{\odot}$ \citep{DonatiGregory2011,DebesJangCondell2013}. TW~Hya is surrounded by a circumstellar disk, with $M_{\rm disk} \gtrsim0.06\,M_{\odot}$ \citep{BerginCleeves2013}, that is viewed almost pole-on \citep[$i\approx7^{\circ}$; ][]{QiHo2004}. Several gaps and features in the disk indicate that planet formation and/or photoevaporation processes are ongoing \citep[e.g.,][and references therein]{HughesWilner2007,AndrewsWilner2016,ErcolanoRosotti2017}. Because of this orientation, the determination of the rotational period of TW~Hya is very controversial. Radial velocity studies provided a periodicity of 3.6\,d \citep{SetiawanHenning2008}, which\ could be interpreted as the stellar rotational in the hypothesis of variations due to photospheric spots \citep{HuelamoFigueira2008}. Conversely, several photometric monitoring campaigns provided different and sparse periodicities \citep[$P_{\rm rot }$ ranging between $\sim1$ and $\sim5$\,d; e.g.,][]{HerbstKoret1988,LawsonCrause2005,SiwakRucinski2011,SiwakRucinski2014}. 

TW~Hya is still actively accreting matter from its disk with a rate that varies between $\sim0.5$ and $\sim2.0\times10^{-9}\,{M_{\odot} \rm yr^{-1}}$ \citep{MuzerolleCalvet2000,CurranArgiroffi2011,DonatiGregory2011}. Accretion is certainly regulated by the stellar magnetic field, since TW~Hya produces a strong field that is characterized by an average surface intensity of $\sim1.5$\,kG, but is capable of reaching values of $\sim3$\,kG in some magnetic spots \citep{JohnsKrullValenti2001,DonatiGregory2011}. In particular \citet{DonatiGregory2011} found that the magnetic configuration is complex and characterized by a strong octupolar component. This, on the one hand, suggests that the magnetosphere is able to disrupt the inner disk at most at distances of $3-4\,R_{\star}$\citep{DonatiGregory2011}, and hints that the geometry of the accretion streams is likely more complicated than that of a pure dipolar configuration.

Observational constraints on the accretion geometry provide a complex scenario. \citet{BatalhaBatalha2002} detected periodic veiling variations and suggested the presence of a hot spot located at low latitudes ($\lesssim20^{\circ}$). Conversely the H$\beta$ line profile, which is characterized by a slightly redshifted absorption feature, points to a hot spot always visible during stellar rotation, but seen through the accretion stream \citep{AlencarBatalha2002}; this scenario is more compatible with high-latitude accretion streams. The different accretion geometries obtained at different epochs, and with different techniques, could be partly reconciled by the spectropolarimetric study presented by \citet{DonatiGregory2011}. Assuming that the observed variations are due to rotational modulation, these authors suggested that on TW~Hya significant accretion can occur both at high and low latitudes, in agreement with the inferred magnetic field configuration. It is worth noting that all the results based on rotational modulation, because of the stellar pole-on view, unknown rotational period, and intrinsic accretion variability, should be considered with caution.

TW~Hya, because it is the nearest CTTS, was targeted several times with {\it Chandra} and {\it XMM} \citep{KastnerHuenemoerder2002,StelzerSchmitt2004,NessSchmitt2005,ArgiroffiMaggio2009,BrickhouseCranmer2010,BrickhouseCranmer2012}. It is X-ray bright, with $L_{\rm X}=1.4\times10^{30}\,{\rm erg\,s^{-1}}$. The emission measure distribution\footnote{The EMD is the distribution in temperature of emission measure, EM, defined as the product of the electron density, hydrogen density, and volume.} (EMD) of its X-ray emitting plasma clearly shows the presence of two main plasma components: a cool ($T\sim3$\,MK) and high-density ($n_{\rm e}\sim (0.6-6.0)\times 10^{12}\,{\rm cm^{-3}}$) plasma, likely due to accretion, and hotter components ($T\gtrsim10$\,MK), reasonably related to coronal activity \citep{KastnerHuenemoerder2002,BrickhouseCranmer2010}. The clear temperature separation of these two components allows us to infer that $\sim75$\% of the total X-ray luminosity is due to the accretion-related plasma. This fraction increases up to 95\% when only cool X-ray emission lines are considered. The chemical composition of the X-ray emitting plasma in TW~Hya is peculiar: it is very rich in neon with respect to other elements. Because the X-ray emitting plasma in TW Hya is related to accretion and hence possibly originates in the inner disk, this plasma could have experienced grain depletion, thereby preferentially maintaining volatile elements, such as neon, that remain in the gas phase \citep{DrakeTesta2005}.

\begin{table*}
\caption{Nominal properties of the {\it Chandra}/HETGS, considering only the positive and negative first-order diffraction spectra.}
\label{tabhetgs}
\begin{center}
\begin{tabular}{ccccccccc}
\hline\hline
        &          &          & \multicolumn{2}{c}{LRF} & & \multicolumn{2}{c}{Wavelength accuracy} &                        \\
\cline{4-5}\cline{7-8}
grating & range    & bin size & $FWHM$ & $\sigma$       & & Absolute       & Relative               &  Eff. Area (@ 10\,\AA) \\
        & (\AA)    & (m\AA)   & (m\AA) & (m\AA)         & & (m\AA)         & (m\AA)                 &  (${\rm cm^2}$)        \\
\hline
HEG     & $1.2-15$ & 2.5      & 12     &  5.1           & &  6             & 1.0                    &  20 \\
MEG     & $1.2-31$ & 5.0      & 23     &  9.8           & & 11             & 2.0                    &  51 \\
\hline
\end{tabular}
\end{center}

\normalsize
\end{table*}

\section{Chandra/HETGS data}
\label{chandraprop}

To search for Doppler shifts in the X-ray emission from TW~Hya, we analyzed the {\it Chandra}/HETGS spectrum collected during the Large Observing Project performed in 2007 (ObsID~7435, 7436, and 7437, PI N.~Brickhouse), and shown in Fig.~\ref{plothetgspec}. Moreover, thanks to the long exposure ($\sim500$\,ks), this observation yielded a very high signal-to-noise ratio ($S/N$) X-ray spectrum rich in emission lines (Fig.~\ref{plothetgspec}). To check for any variability in the measured Doppler shift, we also analyzed the shorter {\it Chandra}/HETGS observation of TW~Hya (exposure time $\sim50$\,ks, ObsID~5, PI C.~Canizares) performed on 2000 July 18, even  though its $S/N$ is significantly lower.

As reported in Sect.~\ref{hetgprop}, looking for Doppler shifts related to plasma motion of $\sim100\,{\rm km\,s^{-1}}$ or lower (depending on the inclination) means pushing the spectral resolution of the {\it Chandra} gratings to the limit. Therefore, to test our methods and to check the wavelength accuracy of the instrument, we also considered other {\it Chandra}/HETGS observations of active stars as a reference (see Sect.~\ref{refspectra}).

All the MEG and HEG data analyzed in this work have been retrieved from the Grating-Data Archive and Catalog \citep{HuenemoerderMitschang2011}.

\begin{figure}
\centering
\includegraphics[width=8.5cm]{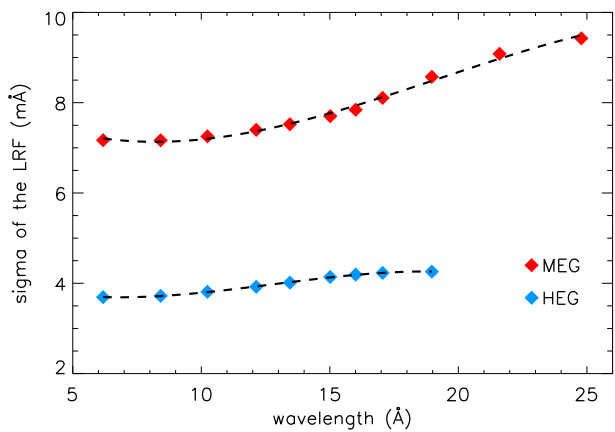}
\caption{Predicted Gaussian $\sigma$ of the LRF of the HEG and MEG gratings, computed at the wavelength corresponding to the most intense emission lines. The black dotted curves indicate the best-fit polynomial function adopted to infer the predicted $\sigma$ at each wavelength.}
\label{plotlinewdt}
\end{figure}

\subsection{Spectral resolution of the {\it Chandra}/HETGS}
\label{hetgprop}

The {\it Chandra}/HETGS currently provides the highest spectral resolving power available for observations of cosmic sources in the X-ray band. This instrument consists of two gratings, the HEG and the MEG, that, when used with the ACIS-S detector, allow collection of two spectra whose nominal properties\footnote{http://cxc.harvard.edu/proposer/POG/html/chap8.html} are reported in Table~\ref{tabhetgs}. To optimize our methods for Doppler shift measurements, and hence to obtain plasma velocities with the highest possible accuracy, we inspected these nominal properties and how they vary with wavelength.

As a first step we inspected the width of the line response function (LRF), as tabulated in the redistribution matrix function of the two gratings, and how it varies within the wavelength range. To this aim we synthesized the spectra corresponding to individual and isolated emission lines, with no intrinsic broadening, at wavelengths corresponding to the strongest lines in the explored range (i.e., $5<\lambda<25$\,\AA). We then fitted these simulated line profiles with a Gaussian function. The corresponding $\sigma$ values obtained from these line fits are plotted in Fig.~\ref{plotlinewdt}. For both HEG and MEG these $\sigma$ are slightly smaller than the corresponding nominal values (see Table~\ref{tabhetgs}), and they smoothly increase for increasing wavelengths.

The precision provided by centroid measurements in individual lines is expected to scale as $\sigma/\sqrt{N}$, with $N$ indicating the number of counts in the line. Therefore, in lines with a few hundreds of counts in the MEG grating, a precision of $\sim10-20\,{\rm km\,s^{-1}}$ can be reached in individual line position.  By combining results from more lines a higher precision can, in principle, be obtained.

\begin{table*}
\caption{Selected sample of stars and their corresponding {\it Chandra}/HETGS observation.}
\label{tab:stellarsample}
\small
\begin{center}
\begin{tabular}{lcccrccrcrc}
\hline\hline
      & \multicolumn{4}{c}{{\it Chandra} Obs. info}                          & & \multicolumn{2}{c}{Stellar properties}                           & & \multicolumn{1}{c}{Earth velocity}                     & ref.$^{c}$ \\
\cline{2-5}\cline{7-8}\cline{10-10}
 Star & ObsID & Obs. start & Obs. stop & \multicolumn{1}{c}{Exp. Time (ks)} & & SpTp & \multicolumn{1}{c}{$v_{\rm phot}^{a} ({\rm km\,s^{-1}})$} & & \multicolumn{1}{c}{$v_{\rm E}^{b} ({\rm km\,s^{-1}})$} &            \\
\hline
     TW Hya (1) &                 7435+7436+7437 & 2007-02-15 & 2007-03-02 & 468.7\hspace{7mm} &  & M0 & $ 12.4$\hspace{5mm} &  & $ 11.8$\hspace{5mm} &  1 \\
     TW Hya (2) &                              5 & 2000-07-18 & 2000-07-18 &  47.7\hspace{7mm} &  & M0 & $ 12.4$\hspace{5mm} &  & $-21.3$\hspace{5mm} &  1 \\
         AU Mic &                             17 & 2000-11-12 & 2000-11-12 &  58.8\hspace{7mm} &  & M1 & $ -4.1$\hspace{5mm} &  & $-28.2$\hspace{5mm} &  2 \\
     EV Lac (1) &                           1885 & 2001-09-19 & 2001-09-20 & 100.0\hspace{7mm} &  & M4.5 & $  0.5$\hspace{5mm} &  & $  2.7$\hspace{5mm} &  2 \\
     EV Lac (2) &                          10679 & 2009-03-13 & 2009-03-14 &  95.6\hspace{7mm} &  & M4.5 & $  0.5$\hspace{5mm} &  & $ -4.8$\hspace{5mm} &  2 \\
         AD Leo &                           2570 & 2002-06-01 & 2002-06-01 &  45.2\hspace{7mm} &  & M4.5 & $ 12.4$\hspace{5mm} &  & $-28.5$\hspace{5mm} &  3 \\
    $\beta$ Cet &                            974 & 2001-06-29 & 2001-07-01 &  86.6\hspace{7mm} &  & K0 & $ 13.3$\hspace{5mm} &  & $ 27.2$\hspace{5mm} &  4 \\
  $\lambda$ And &                            609 & 1999-12-17 & 1999-12-18 &  81.9\hspace{7mm} &  & G8 & $  7.0$\hspace{5mm} &  & $-20.2$\hspace{5mm} &  4 \\
      $\xi$ Uma &                           1894 & 2000-12-28 & 2000-12-29 &  70.9\hspace{7mm} &  & F8.5+G2 & $-18.2$\hspace{5mm} &  & $ 23.7$\hspace{5mm} &  5 \\
         24 Uma &                      2564+3471 & 2002-03-26 & 2002-03-30 &  88.9\hspace{7mm} &  & G9 & $-27.0$\hspace{5mm} &  & $-17.6$\hspace{5mm} &  6 \\
\hline
\end{tabular}
\end{center}
\footnotesize
$^a$ Radial photospheric velocity in the heliocentric reference frame.
$^b$ Radial velocity of the Earth along the line of sight, in the heliocentric reference frame, computed at the time corresponding to the middle between the start and the stop of each observation.
$^c$ References for radial photospheric velocities:  1~\citet{SetiawanHenning2008},  2~\citet{BaileyWhite2012},  3~\citet{NideverMarcy2002},  4~\citet{MassarottiLatham2008},  5~\citet{NordstromMayor2004},  6~\citet{Gontcharov2006}.
\normalsize
\end{table*}

To measure Doppler shifts, it is also fundamental to understand the nominal accuracy of line positions determined by the {\it Chandra}/HETGS gratings, which depends on its wavelength calibration. As reported in Table~\ref{tabhetgs}, the absolute wavelength accuracy is comparable with the line $\sigma$, hence it corresponds to a velocity of $\sim100\,{\rm km\,s^{-1}}$. The accuracy in relative line position within and between observations is instead smaller by a factor of 5. These nominal accuracies of 1.0 and 2.0\,m\AA\, correspond to velocities of 20 and $40\,{\rm km\,s^{-1}}$, respectively, at wavelengths of $\sim15$\,\AA. Therefore, comparing individual line positions between observations, the {\it Chandra}/HETGS allows the tracking of plasma motions down to velocities of a few tens of ${\rm km\,s^{-1}}$. Several studies \citep{BrickhouseDupree2001,ChungDrake2004,IshibashiDewey2006} confirmed this accuracy and proved that the {\it Chandra}/HETGS is very stable in relative wavelength measurements.

These considerations confirm that the velocities expected for the post-shock plasma in CTTS can be detected with the {\it Chandra} gratings, based on relative line positions between different observations. 

\subsection{Reference X-ray spectra}
\label{refspectra}

As reported in the previous section, the highest wavelength accuracy with {\it Chandra}/HETGS data can be obtained comparing line positions between observations. To verify any measured Doppler shift in the X-ray spectrum of TW~Hya, we therefore also checked a selected sample of high $S/N$ spectra collected with the {\it Chandra}/HETGS of active stars without accretion or other expected sources of significant Doppler shifts. In active stars, X-rays are emitted by coronal plasmas, which are located in the outer stellar atmosphere. Therefore it is expected that any Doppler shift in the X-ray spectrum of an active star should yield the same radial velocity as the stellar photosphere. To avoid complications due to variable radial velocity or line broadening, we selected only stars that have negligible radial velocity due to orbital motion and negligible line broadening due to rotation. The selected sample of {\it Chandra}/HETGS observations is reported in Table~\ref{tab:stellarsample}, in which we also list the main characteristics of these active stars together with their known radial velocity\footnote{Throughout the paper we assume that positive radial velocities indicate outward motion (redshift), while negative velocities indicate inward motion (blueshift).}. In this stellar sample, there are also a few multiple systems. In all cases, however, because of the mass ratio and/or because of companion separation, it is known that orbital velocities are negligible ($v\lesssim10\,{\rm km\,s^{-1}}$). 

For stars with multiple {\it Chandra} observations, we considered the HEG and MEG spectra obtained by adding the total exposure time, but only if the different observations were performed approximately at the same epoch (observations 7435, 7436, and 7437 of TW~Hya, and observations 2564 and 3471 of 24~UMa, as also reported in Table~\ref{tab:stellarsample}). This is because, only in those cases, the observations were performed in time intervals small enough to have Earth velocity variations at most of a few ${\rm km\,s^{-1}}$, in the heliocentric reference frame. For the same reason we investigated separately the two observing campaigns of TW~Hya, performed on February 2007 and July 2000 and indicated throughout the paper as TW~Hya~(1) and TW~Hya~(2), and the two observations of EV~Lac, performed on September 2001 and March 2009 and indicated as EV~Lac~(1) and EV~Lac~(2). In Table~\ref{tab:stellarsample} we also report the values of the line-of-sight Earth velocity in the heliocentric reference frame at the epoch of each observation\footnote{Also, we assume positive and negative values for outward and inward motion for the line-of-sight Earth velocity in the heliocentric reference frame, respectively.}.

\section{Doppler shift measurements and plasma velocity}
\label{methods}

We describe in this section the two methods adopted to measure Doppler shifts from the {\it Chandra}/HETGS spectra. We followed two different approaches. The first method is based on measuring the positions of a selected sample of lines, and then averaging these velocities to obtain the global velocity corresponding to each observation. Considering the line formation temperatures, we can adjust this procedure so as to consider only lines produced by plasma components at a few MK. The second method is based on fitting the spectrum with parameterized radiative loss models using the CIAO {\it sherpa} fitting engine and allowing the line-of-sight velocities for cooler accretion-dominated and hotter emission components to vary independently. 

\subsection{Method 1: Individual line positions}
\label{meth1}

Searching for Doppler shift only in individual selected lines allows us to avoid any line blending problem and to monitor only the velocities of plasma components in small temperature ranges.

\begin{table}
\caption{Emission lines selected for Doppler shift measurements.}
\label{tab:linesel}
\normalsize
\begin{center}
\begin{tabular}{lrlcccc}
\hline\hline
 Index$^a$ & \multicolumn{1}{c}{$\lambda^b$} & Elem. & $T_{\rm max}$ & \multicolumn{2}{c}{Flux contrib.$^c$} & line   \\
\cline{5-6}
       & \multicolumn{1}{c}{($\AA$)}     &       & (MK)          & Accr. & Cor.                          & subset$^d$ \\
\hline
 1a &   6.1804 &  \ion{Si}{XIV} & 15.8 & 0.00 & 1.00 & h \\
 1b &   6.1858 &  \ion{Si}{XIV} & 15.8 & 0.00 & 1.00 & h \\
 2  &   6.6479 &  \ion{Si}{XIII} & 10.0 & 0.01 & 0.99 & h \\
 3  &   6.7403 &  \ion{Si}{XIII} & 10.0 & 0.02 & 0.98 & h \\
 4a &   8.4192 &  \ion{Mg}{XII} & 10.0 & 0.00 & 1.00 & h \\
 4b &   8.4246 &  \ion{Mg}{XII} & 10.0 & 0.00 & 1.00 & h \\
 5  &   9.1687 &  \ion{Mg}{XI} &  6.3 & 0.31 & 0.69 & h \\
 6a &   9.7080 &  \ion{Ne}{X} &  6.3 & 0.07 & 0.93 & h \\
 6b &   9.7085 &  \ion{Ne}{X} &  6.3 & 0.07 & 0.93 & h \\
 7a &  10.2385 &  \ion{Ne}{X} &  6.3 & 0.09 & 0.91 & h \\
 7b &  10.2396 &  \ion{Ne}{X} &  6.3 & 0.09 & 0.91 & h \\
 8  &  13.6990 &  \ion{Ne}{IX} &  4.0 & 0.96 & 0.04 & c \\
 9  &  15.0140 &  \ion{Fe}{XVII} &  5.0 & 0.73 & 0.27 & h \\
10a &  16.0055 &  \ion{O}{VIII} &  3.2 & 0.92 & 0.08 & c \\
10b &  16.0067 &  \ion{O}{VIII} &  3.2 & 0.92 & 0.08 & c \\
11  &  16.7800 &  \ion{Fe}{XVII} &  5.0 & 0.77 & 0.23 & h \\
12  &  17.0510 &  \ion{Fe}{XVII} &  5.0 & 0.76 & 0.24 & h \\
13  &  17.0960 &  \ion{Fe}{XVII} &  5.0 & 0.78 & 0.22 & h \\
14a &  18.9671 &  \ion{O}{VIII} &  3.2 & 0.94 & 0.06 & c \\
14b &  18.9725 &  \ion{O}{VIII} &  3.2 & 0.94 & 0.06 & c \\
15  &  21.6015 &  \ion{O}{VII} &  2.0 & 1.00 & 0.00 & c \\
\hline
\end{tabular}
\end{center}
\footnotesize
$^a$ Lines indicated with the same index number, but different letters, are the two components of H-like resonance doublets.
$^b$ APED predicted wavelengths.
$^c$ Fractional contributions to individual line fluxes due to accretion and corona components, on the basis of the {\it model C} obtained by \citet{BrickhouseCranmer2010}.
$^d$ "c" and "h" indicate whether the line has been included in the cool or hot line subsample.
\normalsize
\end{table}

\subsubsection{Line selection}
\label{linesel}

Among the tens of emission lines typically observed in the TW~Hya X-ray spectrum (Fig.~\ref{plothetgspec}), we first selected a sample of emission lines, including only strong and isolated lines. The selected lines are listed in Table~\ref{tab:linesel}, where we also report their rest wavelengths, as predicted by the APED atomic databases \citep{SmithBrickhouse2001}, and their maximum formation temperature.

Although we are only interested in single lines, we also included some resonance doublets of H-like ions in our sample (indicated in Table~\ref{tab:linesel} with the same index number, but different letters). Considering the spectral resolution of the HEG and MEG gratings, the two components of each doublet are not resolved, but their wavelength difference is not negligible. This is not an issue for the line position determination since the relative position and relative intensity (exactly equal to 2:1 in the optically thin emission regime) of the two contributions are known. Therefore, even if these doublets are not strictly isolated single lines, the total doublet profile is perfectly known.

\begin{figure}
\centering
\includegraphics[width=8.5cm]{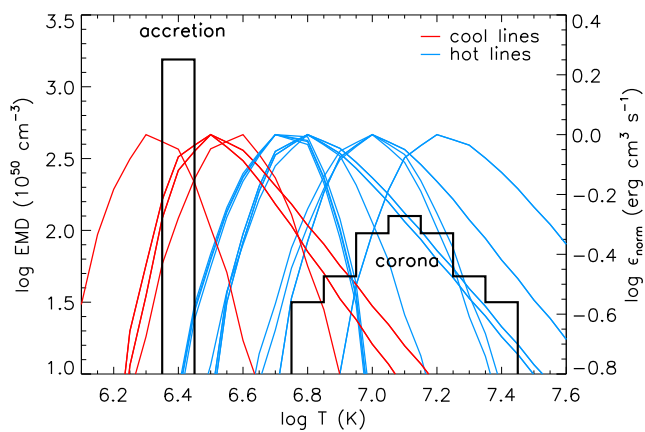}
\caption{EMD (in black) of the X-ray emitting plasma in TW~Hya, composed by an accretion component and a coronal component \citep[{\it model C} derived by][]{BrickhouseCranmer2010}. Red and blue curves indicate the normalized emissivity function $\epsilon_{norm}$ of the cool and hot selected lines.}
\label{fig:checklineorigin}
\end{figure}

Accretion shocks in CTTS can generate plasma whose temperatures are at most a few MK. A Doppler shift with respect to the stellar photosphere is therefore expected only for the X-ray lines produced at these temperatures. The maximum formation temperature $T_{\rm max}$ of the selected lines ranges from 2 up to 16\,MK. We considered the EMD of TW~Hya \citep[{\it model C} from][]{BrickhouseCranmer2010} to better constrain the origin of each selected line;  this EMD is characterized by a sharp peak at $T\approx2.5$\,MK ascribed to the accretion shock, and by a large peak centered at $T\approx13$\,MK ascribed to coronal plasma (Fig.~\ref{fig:checklineorigin}). This EMD, together with the emissivity functions $\epsilon(T)$ of each selected line (also shown in Fig.~\ref{fig:checklineorigin}, normalized to their maximum), allowed us to compute the expected contributions of accretion and coronal components to the flux of each line. These relative contributions are reported in Table~\ref{tab:linesel}. We therefore selected the subset of cool lines (labeled with ``c'' in Table~\ref{tab:linesel} and with red curves in Fig.~\ref{fig:checklineorigin}). All these cool lines have $T_{\rm max} < 4.5$\,MK and any coronal contribution to their flux is expected to be less than 10\%. Therefore these lines should allow us to probe the velocity of the plasma heated in the accretion shock. The remaining lines, indicated as hot lines (labeled with ``h'' in Table~\ref{tab:linesel} and with blue curves in Fig.~\ref{fig:checklineorigin}) have maximum formation temperatures higher than 4.5\,MK, and a non-negligible coronal contribution to their flux. Some of these hot lines, especially the \ion{Fe}{xvii} lines, have significant contributions from both accretion and coronal plasma components. Their Doppler shifts might therefore be an average between these two plasma components.

Considering the full line set we checked the effect of possible blends due to other weak emission lines, which were not detectable individually but could skew the line position estimation. To this aim, we checked that the expected contribution of weak and unresolved lines in the wavelength regions of the lines selected for analysis was lower than $\sim5$\%. Therefore, blending problems due to unresolved lines are expected to be negligible in the determination of the centroid position of each line.

\begin{figure}
\centering
\includegraphics[width=8.15cm]{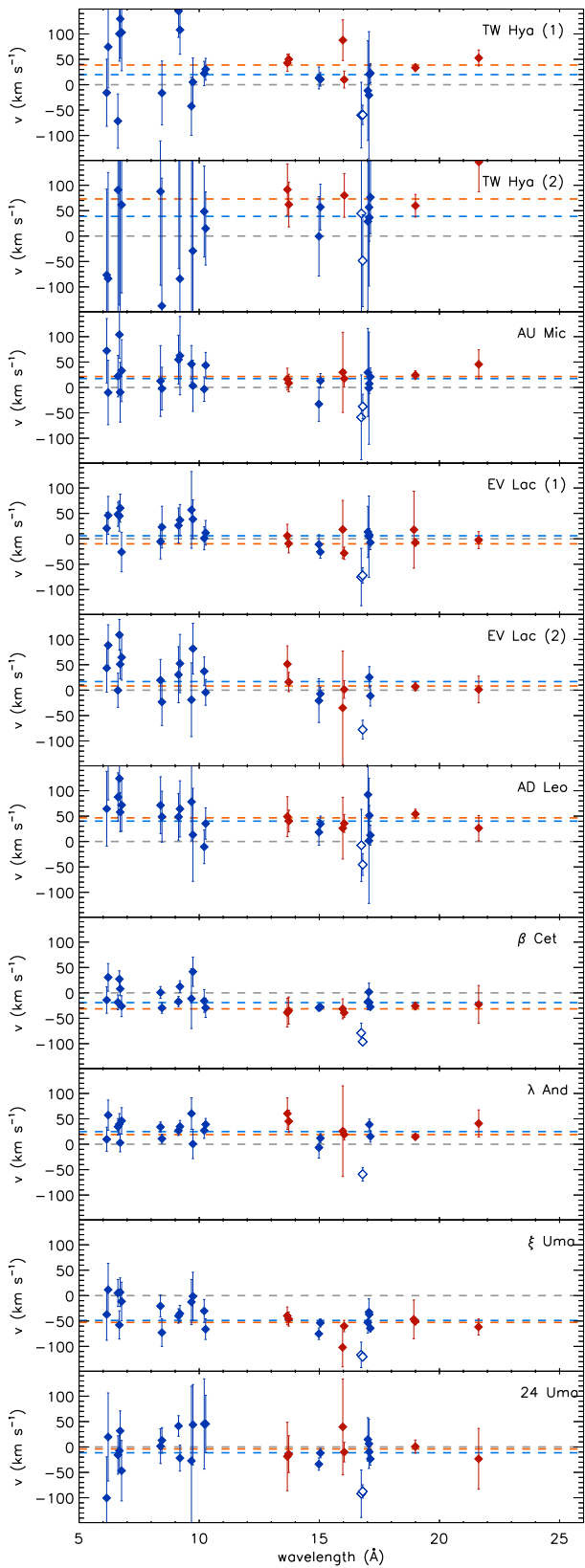}
\caption{Velocity obtained from each fitted line, for each inspected spectrum, with respect to the {\it Chandra} satellite reference frame. The red and blue diamonds indicate lines included in the cool and hot line sample, respectively, and errors refer to the 1\,$\sigma$ level. The red and blue dashed lines correspond to the average velocities, $v_{\rm Xc}^{ }$ and $v_{\rm Xh}^{ }$, obtained for the cool and hot line sample. The velocity measured for the \ion{Fe}{xvii} line at 16.78\,\AA, not included in the average velocity computation, is indicated with an open diamond. To clarify, we included a small wavelength offset in those points corresponding to the same line fitted in the two gratings.}
\label{fig:plotlinevel}
\end{figure}

\subsubsection{Line position measurement}
\label{linepos}

We determined the position of the selected emission lines by fitting separately each line profile in the HEG and MEG spectra, obtained by adding positive and negative first order diffraction spectra. We performed the fit in a small wavelength interval of $\sim0.1-0.2$\,\AA, centered on the predicted line wavelength, for each grating and for each
line. For some observations, some lines in the HEG grating are almost undetected. We therefore performed the line fit only when a minimum number of 5 counts were gathered\footnote{The value of 5 cts threshold, set for very weak lines, does not significantly affect our procedure. The positions of marginally detected lines are in any case characterized by large uncertainty, hence these lines do not significantly affect the final determination of the plasma radial velocity obtained by a weighted average (see Sect.~\ref{avrgvel}).}  in the selected wavelength
interval. As a best-fit function we assumed a Gaussian profile plus a constant, the latter serving to represent the continuum level. In the cases of the H-like resonance doublets, we fitted the observed profiles with two Gaussians, whose displacement and relative intensity are fixed to the known theoretical values. In all cases, to minimize the errors on the measurements of the line positions, we performed the fits fixing the $\sigma$ of the Gaussian profile  to the value predicted by the LRF and reported in Sect.~\ref{hetgprop}. The observed and fitted line profiles are shown in Figs.~\ref{fig:fitlines1} and \ref{fig:fitlines2}. 

\subsubsection{Determination of the radial velocity of the X-ray emitting plasma}
\label{avrgvel}

For each fitted line, we determined the velocity corresponding to the wavelength shift with respect to the predicted wavelength rest value. The velocities obtained are reported in Table~\ref{tab:linewvl} and plotted in Fig.~\ref{fig:plotlinevel}. These velocities are computed with respect to the {\it Chandra} satellite reference frame.

Considering all the {\it Chandra}/HETGS spectra adopted as a reference, the velocities provided from different lines do not show any systematic pattern across the explored wavelength range.  The only exception to this lack of pattern is the \ion{Fe}{xvii} line at 16.78\,\AA~ (denoted with an open symbol in Fig.~\ref{fig:plotlinevel}); in all the observations this line is always significantly and systematically blueshifted, if compared to the other lines of the same observation. We therefore excluded this line from our subsequent analysis, assuming that there is probably some issue in its predicted rest wavelength.

Since the other lines do not show any systematic trend, we assumed that the individual line velocities are random fluctuations of the X-ray plasma velocity. Therefore, for each observation we inferred the plasma velocity by computing the weighted average of the velocities obtained from individual lines, considering separately the cool and hot line subsamples, as reported in Sect.~\ref{linesel} and as indicated in Table~\ref{tab:linesel}. These cool and hot average velocities\footnote{These are always the velocities with respect to the {\it Chandra} satellite reference frame.}, $v_{\rm Xc}^{ }$ and $v_{\rm Xh}^{ }$ , are reported in Table~\ref{tab:linewvl}, and plotted as horizontal dashed lines in Fig.~\ref{fig:plotlinevel}.

We performed further checks to investigate whether the final average velocities, $v_{\rm Xc}^{ }$ and $v_{\rm Xh}^{ }$, depend on including or not the HEG spectra (that are of higher spectral resolution but characterized by significantly lower $S/N$ and provide significantly larger uncertainties in the line velocities), and/or depend on considering separately the $+1$ and $-1$ diffraction orders. We found that the average velocities, $v_{\rm Xc}^{ }$ and $v_{\rm Xh}^{ }$ , do not change significantly when computed with various approaches.

\subsection{Method 2: Spectral model parameter estimation}
\label{meth2}

The second method we used for Doppler shift measurements consists in modeling the whole X-ray spectrum at once. This approach offers the advantage, at least in principle, of implicitly including all of the velocity information contained in the data into the analysis.

We examined the long TW~Hya observation, obtained in 2007, indicated here as TW~Hya~(1), and composed by three individual exposures of $\approx$150\,ks each (ObsIDs 7435, 7436 and 7437). We modeled them both separately and combined.

\begin{table*}
\caption{Spectral model parameter estimation for TW Hya.}
\small
\label{tab:sherpa}
\begin{center}
\begin{tabular}{llccccc}
\hline\hline
      &                                           & \multicolumn{5}{c}{Observation} \\
\cline{3-7}
Comp. & Parameter                                 & TW Hya (1)$^a$              & TW Hya (7435)             & TW Hya (7436)             & TW Hya (7437)             & TW Hya (7435+7436)         \\ 
\hline
      & $N_{\rm H}^b$ ($10^{20}\,{\rm cm^{-2}})$  & $10$                        & $10$                      & $10$                      & $10$                      & $10$                       \\ 
\hline
Cool  & $T$$^b$ (MK)                              & $2.5$                       & $2.5$                     & $2.5$                     & $2.5$                     & $2.5$                      \\ 
      & Ne$^b$ (Ne$_{\odot}$)                     & $1.2$                       & $1.2$                     & $1.2$                     & $1.2$                     & $1.2$                      \\ 
      & Fe$^b$ (Fe$_{\odot}$)                     & $0.2$                       & $0.2$                     & $0.2$                     & $0.2$                     & $0.2$                      \\ 
      & $v_{\rm Xc}$ (${\rm km\,s^{-1}}$)         & $44 \pm 2$                  & $40 \pm 4$                & $47 \pm 4$                & $44 \pm 4$                & $44 \pm 3$                 \\ 
      & Norm$^c$ ($10^4$)                         & $76.7 \pm 1.3$              & $83.6 \pm 2.1$            & $102 \pm 2.3$             & $87.6 \pm 2.2$            & $81.3 \pm 1.5$             \\ 
 \hline
Hot   & $T$ (MK)                                  & $11.59 \pm 0.15$            & $11.13 \pm 0.27$          & $11.54 \pm 0.24$          & $12.02 \pm 0.25$          & $11.36 \pm 0.18$           \\ 
      & Ne (Ne$_{\odot}$)                         & $1.38 \pm 0.06$             & $1.38 \pm 0.12$           & $1.38 \pm 0.11$           & $1.37 \pm 0.10$           & $1.39 \pm 0.08$            \\ 
      & Fe (Fe$_{\odot}$)                         & $0.086 \pm 0.004$           & $0.084 \pm 0.008$         & $0.085 \pm 0.007$         & $0.087 \pm 0.007$         & $0.085 \pm 0.005$          \\ 
      & $v_{\rm Xh}$ (${\rm km\,s^{-1}}$)         & $25 \pm 6 $                 & $9.4 \pm 10$              & $25 \pm 9$                & $37 \pm 9$                & $19 \pm 7$                 \\ 
      & Norm$^c$ ($10^4$)                         & $1.51 \pm .04$              & $1.46 \pm 0.07$           & $1.88 \pm 0.07$           & $2.01 \pm 0.08$           & $1.46 \pm 0.05$            \\ 
\hline
      & 7436 Norm$^d$                             & $1.28 \pm 0.02$             & \dots                     & \dots                     & \dots                     & $1.28 \pm 0.02$            \\ 
      & 7437 Norm$^d$                             & $1.28 \pm 0.02$             & \dots                     & \dots                     & \dots                     & \dots                      \\ 
\hline
\end{tabular}
\end{center}
All the errors refer to the 1\,$\sigma$ level. $^a$ This column reports the results obtained considering the combination of the ObsID~7435+7436+7437, we labeled it as TW~Hya~(1) in agreement with the rest of the work.
$^b$ Parameter fixed.
$^c$ Normalization parameter in units of $10^{-14} / (4 \pi d^2) \int n_{\rm e} n_{\rm H} \, {\rm d}V,$ where $d$ is the distance to the source (cm), $n_{\rm e}$ is the plasma electron density (${\rm cm^{-3}}$), and $n_{\rm H}$ is the plasma hydrogen density (${\rm cm^{-3}}$).
$^d$ Global normalization constant applied multiplicatively to the whole model (both hot and cool components) in simultaneous fit.
\normalsize
\end{table*}

\begin{table*}
\caption{Radial velocities of the cool and hot X-ray emitting plasmas in the {\it Chandra} ($v_{\rm Xc}^{ }$ and $v_{\rm Xh}^{ }$) and in the Stellar ($v_{\rm Xc}^{\star}$ and $v_{\rm Xh}^{\star}$) reference frames.}
\label{tab:vrad}
\normalsize
\begin{center}
\begin{tabular}{lrrrrrr}
\hline\hline
 Star & \multicolumn{1}{c}{$v_{\rm E}^a$} & \multicolumn{1}{c}{$v_{\rm phot}^b$} & \multicolumn{1}{c}{$v_{\rm Xc}^{ }$}     & \multicolumn{1}{c}{$v_{\rm Xh}^{ }$}     & \multicolumn{1}{c}{$v_{\rm Xc}^{\star}$}     & \multicolumn{1}{c}{$v_{\rm Xh}^{\star}$}  \\
      & \multicolumn{2}{c}{(Heliocentric ref. frame)}                            & \multicolumn{2}{c}{({\it Chandra} ref. frame)}                                      & \multicolumn{2}{c}{(Stellar ref. frame)} \\
\hline
\multicolumn{7}{c}{Method 1: individual line positions} \\
\hline
TW Hya (1) &  11.8 &  12.4\hspace{8mm} & $ 38.9 \pm   5.1$ & $ 19.9 \pm   6.9$ & $ 38.3 \pm   5.1$ & $ 19.3 \pm   6.9$ \\
TW Hya (2) & -21.3 &  12.4\hspace{8mm} & $ 73.1 \pm  16.4$ & $ 38.8 \pm  23.2$ & $ 39.3 \pm  16.4$ & $  5.1 \pm  23.2$ \\
AU Mic & -28.2 &  -4.1\hspace{8mm} & $ 21.5 \pm   6.5$ & $ 17.4 \pm   7.1$ & $ -2.6 \pm   6.5$ & $ -6.7 \pm   7.1$ \\
EV Lac (1) &   2.7 &   0.5\hspace{8mm} & $ -9.9 \pm   5.3$ & $  5.9 \pm   5.3$ & $ -7.8 \pm   5.3$ & $  8.1 \pm   5.3$ \\
EV Lac (2) &  -4.8 &   0.5\hspace{8mm} & $  8.2 \pm   6.6$ & $ 16.9 \pm   6.9$ & $  2.8 \pm   6.6$ & $ 11.5 \pm   6.9$ \\
AD Leo & -28.5 &  12.4\hspace{8mm} & $ 46.6 \pm   7.2$ & $ 40.1 \pm   7.6$ & $  5.7 \pm   7.2$ & $ -0.8 \pm   7.6$ \\
$\beta$ Cet &  27.2 &  13.3\hspace{8mm} & $-31.5 \pm   5.4$ & $-19.1 \pm   2.2$ & $-17.6 \pm   5.4$ & $ -5.2 \pm   2.2$ \\
$\lambda$ And & -20.2 &   7.0\hspace{8mm} & $ 18.9 \pm   4.7$ & $ 24.8 \pm   3.2$ & $ -8.3 \pm   4.7$ & $ -2.4 \pm   3.2$ \\
$\xi$ Uma &  23.7 & -18.2\hspace{8mm} & $-52.7 \pm   4.6$ & $-48.9 \pm   3.4$ & $-10.8 \pm   4.6$ & $ -7.0 \pm   3.4$ \\
24 Uma & -17.6 & -27.0\hspace{8mm} & $ -3.9 \pm   9.9$ & $-11.4 \pm   4.7$ & $  5.5 \pm   9.9$ & $ -2.0 \pm   4.7$ \\
\hline
\multicolumn{7}{c}{Method 2: spectral model parameter estimation} \\
\hline
TW Hya (1) &  11.8 &  12.4\hspace{8mm} & $ 44.0 \pm   2.0$ & $ 25.0 \pm   6.0$ & $ 43.4 \pm   2.0$ & $ 24.4 \pm   6.0$ \\
\hline
\end{tabular}
\end{center}
\footnotesize
All the velocities are in ${\rm km\,s^{-1}}$. All the errors refer to the 1 $\sigma$ level. $^a$ Radial velocity of the Earth, computed at the time corresponding to the middle between the start and the stop of each observation.
$^b$ Radial photospheric velocity of each star.
\normalsize
\end{table*}

We considered HEG and MEG spectra in the wavelength ranges of 1.5$-$18\,\AA\, and 2$-$26\,\AA, respectively. We used the {\it sherpa} fitting engine in CIAO version 4.9 with APEC models, and employed the {\it model C} coronal and accretion plasma EMD described by \citet{BrickhouseCranmer2010}, reported here in Fig.~\ref{fig:checklineorigin}. This EMD is  characterized by a fixed temperature component, at 2.5\,MK, representing the post-shock plasma, and a hotter component (a Gaussian-broadened $\delta$ function) centered on 13\,MK. One difference with respect to the analysis of \citet{BrickhouseCranmer2010} is that we allowed the centroid temperature of this latter component to vary during spectral fitting. We also allowed the Ne abundance in this component to vary independently of other metals. As also noted by \citet{BrickhouseCranmer2010}, the abundances of metals in the accreting gas and absorption due to interstellar and ambient gas are not independently well constrained by the data. The absorption due to ambient gas is also probably very complex and non-uniform. Since here we are primarily concerned with understanding Doppler shifts rather than in building an accurate picture of the coronal and accretion plasma characteristics, we fixed the absorption component $N_{\rm H}$ to the single value of $10^{21}\,{\rm cm^{-2}}$, as adopted by \citet{BrickhouseCranmer2010}. Additionally, we fixed the Ne and other metal abundances in the accretion component at 1.2 and 0.2 times the solar values of \citet{AndersGrevesse1989}, respectively, following \citet{DrakeTesta2005} and also guided by the \citet{BrickhouseCranmer2010} results. No attempt was made to match the ratios of the density sensitive He-like ions that were first pointed out, by \citet{KastnerHuenemoerder2002}
and \citet{StelzerSchmitt2004}, as showing higher densities in TW~Hya than in non-accreting magnetically active stars.

We list in Table~\ref{tab:sherpa} the model fit results, together with the velocities of cool and hot plasma components, reporting both the parameters obtained from combined or individual observation analysis. The individual model component normalization parameter (``Norm'' in Table~\ref{tab:sherpa}) depends on plasma emission measure and source distance. In the case of simultaneous modeling of different observations, independent multiplicative constants were introduced to allow the global normalization of the model to vary among the different observation segments.

Considering the results obtained from individual observations of TW~Hya, we found a slightly different velocity for the hot component of ObsID~7437 than for the other two observations. We therefore also performed simultaneous parameter estimation for just ObsID~7435 and 7436 spectra, excluding ObsID~7437, as shown in Table~\ref{tab:sherpa}.

\subsection{Velocity reference frame}

Both the methods applied provide the bulk velocities of the cool and hot plasma components, $v_{\rm Xc}^{ }$ and $v_{\rm Xh}^{ }$, averaged over the whole exposure time of the inspected observations and computed with respect to the {\it Chandra} satellite reference frame. During its orbit, the satellite reaches velocities of at most $\sim1-2\,{\rm km\,s^{-1}}$ with respect to the Earth. Such velocities are negligible with respect to both the errors on the average velocities computed ($\sim5-10\,{\rm km\,s^{-1}}$) and the velocities that we aim to measure (tens of ${\rm km\,s^{-1}}$). We therefore assume that the velocity measured with respect to the satellite reference frame is the same as that with respect to the Earth. Therefore, taking into account the line-of-sight Earth velocity in the heliocentric reference frame at the epoch of each observation $v_{\rm E}^{ }$ (listed in Table~\ref{tab:stellarsample}), the X-ray plasma velocity in the stellar reference frame, $v_{\rm X}^{\star}$, can be computed as $v_{\rm X}^{ }+v_{\rm E}^{ }-v_{\rm phot}^{ }$. All these velocities are listed in Table~\ref{tab:vrad}.

\begin{figure*}
\centering
\includegraphics[width=16cm]{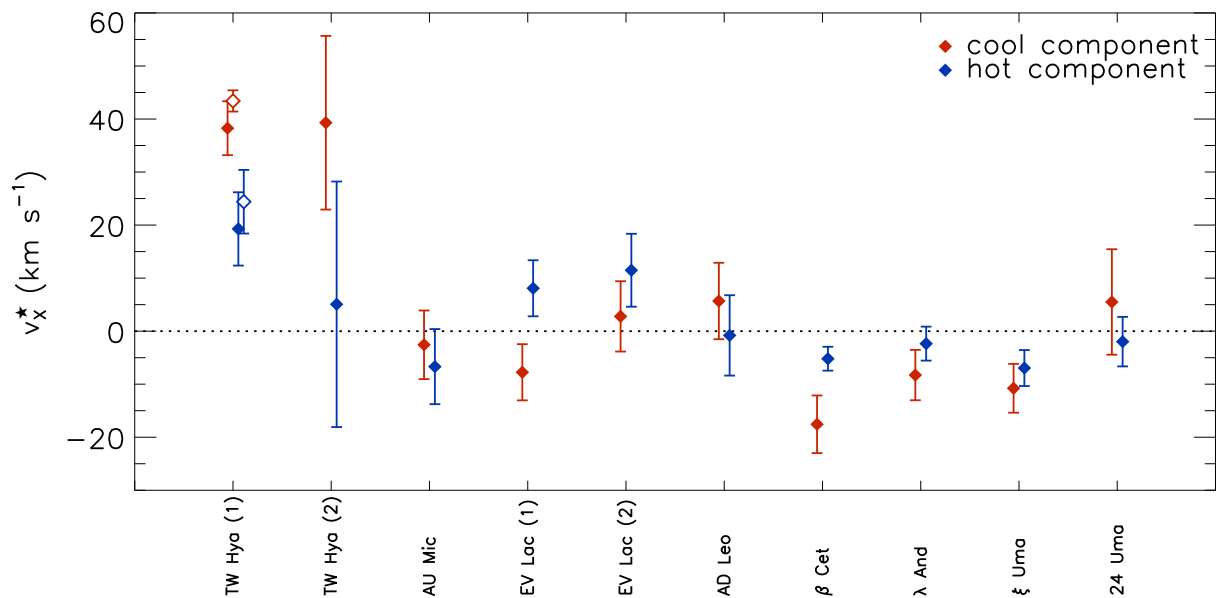}
\caption{Cool and hot X-ray emitting plasma velocities $v_{\rm X}^{\star}$ in the stellar reference frame with errors at the $1\sigma$ level. Filled symbols indicate velocities obtained with method~1 (individual line positions), open symbols indicate velocities obtained with method~2 (spectral model parameter estimation).}
\label{fig:vxvsvphot}
\end{figure*}

\section{Results}
\label{results}

We show in Fig.~\ref{fig:vxvsvphot} the cool and hot plasma velocities for each star, $v_{\rm Xc}^{\star}$ and $v_{\rm Xh}^{\star}$, with respect to the star itself. The first conclusion is that the two methods applied to TW~Hya~(1) to measure $v_{\rm Xc}^{\star}$ and $v_{\rm Xh}^{\star}$ provide compatible results, indicating that the measured velocities do not depend on the methods. In the subsequent discussion, we refer to the $v_{\rm Xc}^{\star}$ and $v_{\rm Xh}^{\star}$ values obtained with method~1, although we stress that velocities obtained with methods~1 and 2 are in excellent agreement.

Inspecting the results obtained for the set of reference stars, where the X-ray emission originates only from coronal plasma, we found that both the cool and hot plasma components provide velocities compatible with zero. The only marginal exception is the cool plasma associated with $\beta$~Cet, which shows a blueshifted velocity with respect to the stellar reference frame of $-17.6\pm 5.4\,{\rm km\,s^{-1}}$, significant at $3.2\sigma$ level. However, considering the entire set of reference stars, their cool plasma velocities are marginally, but systematically, blueshifted with respect to the photosphere, with a mean value of $\sim -4\pm3\,{\rm km\,s^{-1}}$. Taking into account this small offset would also reconcile also the $\beta$~Cet case. We can therefore conclude that the X-ray emitting plasma in purely coronal sources displays on average the same velocity as the underlying stellar photosphere. This finding, on the one hand, confirms the reliability of the adopted procedures to measure the X-ray emitting plasma velocity and, on the other hand, verifies the accuracy of the wavelength calibration of the {\it Chandra} gratings.

TW~Hya clearly shows a different behavior. For the long exposure observation (labeled as TW~Hya~(1)) we found with both our methods that the cool plasma component is significantly redshifted with respect to the stellar photosphere. Method~1 provides a receding velocity of this cool plasma component of $38.3\pm5.1\,{\rm km\,s^{-1}}$. A perfectly compatible value, $42.4\pm2.0\,{\rm km\,s^{-1}}$, comes from method~2.

The TW~Hya~(1) observation is composed of three observing segments, of $\sim150$\,ks each, spread over a time period of a few days (see Table~\ref{tab:stellarsample}). The receding velocity of the cool plasma component does not change significantly considering individual observing segments (see Table~\ref{tab:sherpa}). Moreover, we also found the same receding velocity for the short exposure observation, TW~Hya~(2), performed a few years before (despite the lower significance owing to a lower $S/N$ ratio of this spectrum).

The velocity of the hot plasma component on TW~Hya, determined with both methods~1 or 2, ($19.3\pm 6.9$ and $24.4\pm 6.0$, respectively) is only marginally consistent with zero velocity offset. This could reflect the adopted velocity determination method, which is based on line position. In fact, as already argued in Sect.~\ref{linesel}, the velocity of the hot plasma component is likely an average between both the accretion and coronal components. To confirm this assessment, we simulated synthetic {\it Chandra}/HETGS spectra of TW~Hya assuming that its X-ray emission is due to an accretion and a coronal component \citep[{\it model C}, presented by ][and shown in Fig.~\ref{fig:checklineorigin}]{BrickhouseCranmer2010}, and supposing that the accretion component is receding with respect to the photosphere, while the coronal plasma is fixed. If we apply the same procedure to the total synthetic spectrum that we applied to the real TW~Hya spectrum (method~1, Sect.~\ref{meth1}), the hot plasma component appears slightly redshifted with respect to the assumed zero velocity of the corona. As predicted, this happens because several emission lines, and almost all the strong \ion{Fe}{xvii} lines, have significant contributions from both accretion and coronal components, and hence show a global redshift that is an average between the two assumed input velocities\footnote{The LRF of HEG and MEG gratings does not allow us to resolve the two velocity components in individual line profiles.}. Nevertheless, it could be that the EMD of TW~Hya is not actually composed of two well-separated components (such as those taken as a reference in this work, Fig.~\ref{fig:checklineorigin}), but could also have non-negligible amounts of plasma at $T\sim4-5$\,MK \citep[i.e., $\log T\sim6.6-6.7$; as found by][]{ArgiroffiMaggio2009} that comes from accretion and hence also produces, in the whole X-ray spectrum, redshifted features at temperatures higher than those predicted by the EMD model assumed here.

In the following discussion, focused on the redshift observed for the cool plasma associated with TW~Hya, we interpret this redshift as due to a dominant cool plasma component moving as a whole with the observed velocity. We inspected the profiles of the strongest emission lines in the TW~Hya~(1) HEG spectrum, considering separately +1 and -1 diffraction orders, to validate
this hypothesis of a single velocity component. We fitted these lines with the $\sigma$ left as a free parameter, and we verified that the best-fit values for $\sigma$ were in all cases compatible with the values predicted by the LRF. Therefore in the TW~Hya~(1) line profile there is no evidence of line broadening, and hence of the presence of a distribution of velocities. Considering the intrinsic spectral resolution of the {\it Chandra} gratings, however, we cannot rule out that the observed redshift could come from a mixture of plasma components receding with velocities differing by a few tens of ${\rm km\,s^{-1}}$.

\section{Discussion}
\label{discussion}

Inspecting the X-ray emission from TW~Hya, we detected a redshift, with respect to the photosphere, in the plasma components at $\sim2-4$\,MK. This result provides important constraints on the origin of this plasma component and on the geometry of the accretion streams on TW~Hya.

\subsection{Cool plasma origin in CTTS}

The nature of the cool, high-density plasma component observed in CTTS has been the subject of considerable debate.
On the one hand, the high density ($n_{\rm e}\sim10^{12}\,{\rm cm^{-3}}$) and cool temperature  ($T\sim1-4\,MK$) of this plasma have been the main indications that this material is heated in the accretion shock \citep{KastnerHuenemoerder2002,StelzerSchmitt2004,SchmittRobrade2005,GuntherLiefke2006,ArgiroffiMaggio2007,HuenemoerderKastner2007,ArgiroffiFlaccomio2011}. This scenario was then supported by further constraints: the cool plasma EMD is sharply peaked around $T\sim3$\,MK \citep{BrickhouseCranmer2010} as predicted by HD models \citep{SaccoArgiroffi2008,SaccoOrlando2010}; some strong emission lines emitted by this cool plasma component show a non-negligible optical depth \citep{ArgiroffiMaggio2009} and display rotational modulation \citep{ArgiroffiMaggio2012}. On the other hand, other aspects do not fit easily into this scenario. \citet{GuedelSkinner2007} did not detect any high-density plasma component from the CTTS T~Tauri. \citet{GuedelTelleschi2007} noted that the soft X-ray excess shown by CTTS, with respect to non-accreting young stars, does not correlate with UV accretion indicators. Moreover, \citet{BrickhouseCranmer2010} found that both the density versus temperature pattern (characterized by increasing density for increasing temperature) and the volumes of the cool plasma components of TW~Hya are not reconcilable with the hypothesis of a plasma heated in the accretion shock. Based on the above results, it was suggested that, in addition to plasma heated in the accretion shock, the cool plasma in CTTS could also be composed of coronal structures that are modified, cooled, or fed by accreted material \citep{GuedelTelleschi2007,BrickhouseCranmer2010,SchneiderGunther2017}.

One-dimensional HD modeling of the accretion shock region indicates that on average the hot post-shock plasma is indeed characterized by high-density and relatively cool temperature compared with the X-ray emitting corona, similar to that observed \citep{SaccoArgiroffi2008,SaccoOrlando2010}. Detailed two-dimensional MHD models show that the hot post-shock region can be composed of independent fibrils, and, because of time-dependent thermal instabilities, can have a composite density versus temperature structure \citep{BonitoOrlando2014}.  Moreover, when the effect of local absorption due to the pre-shock material is taken into account, the apparent density versus temperature structure inferred from the emerging X-ray spectrum shows a pattern analogous to that observed \citep{BonitoOrlando2014}.  Two-dimensional MHD models \citep{OrlandoSacco2010} show that any flow of accreted material from the flux tube into the surrounding corona is prevented, unless the local magnetic field is weak, i.e., even a moderate magnetic field of $\sim200$\,G provides a plasma $\beta\sim1$ for a plasma density of $\sim10^{13}\,{\rm cm^{-3}}$. Typical values of surface magnetic fields of CTTS are significantly stronger \citep[$B\sim1-3$\,kG; e.g.,][and references therein]{JohnsKrull2007}, in agreement with magnetospheric accretion that requires accretion tubes to have strong magnetic fields to be able to load and guide inner disk matter.

One of the most important predictions for the plasma heated in the accretion shock, confirmed by detailed HD and MHD simulations, is that it should have a downward bulk velocity within the stellar atmosphere \citep{SaccoArgiroffi2008,SaccoOrlando2010,OrlandoSacco2010,OrlandoBonito2013,BonitoOrlando2014}. Therefore, depending on the inclination under which the shock region is observed, X-ray lines emitted by post-shock plasma are expected to be redshifted. We show in this work that the X-ray emission lines from the cool plasma associated with TW~Hya are indeed redshifted with respect to the stellar photosphere, providing an average receding velocity of $38.3\pm5.1\,{\rm km\,s^{-1}}$. The cool plasma then has an inward motion in the stellar atmosphere, which is in agreement with predictions based on the accretion shock scenario. This is further strong evidence that the plasma at a few MK on TW~Hya, responsible for the soft X-ray emission, is indeed located in the post-shock region and not in accretion-fed coronal loops. This holds not only for the cool plasma as a whole, but also for individual plasma components at low temperatures. We in fact also detected a significant redshift from individual cool lines (see Table~\ref{tab:linewvl} and Fig.~\ref{fig:plotlinevel}). In particular, all the cool plasma components at $\sim2$, $\sim3$, and $\sim4$\,MK associated with TW~Hya, probed by the emission lines of \ion{O}{vii}, \ion{O}{viii}, and \ion{Ne}{ix}, move with similar inward velocities. On TW~Hya, all the cool plasma components responsible for the soft X-ray emission then appear to be located in the post-shock region.

\subsection{Constraints on the accretion geometry from X-ray Doppler shift}

We observed that the soft X-ray emission from TW~Hya is significantly redshifted, $v_{\rm Xc}^{\star}=38.3\pm5.1\,{\rm km\,s^{-1}}$, with respect to the photosphere. The observed radial velocity of the cool plasma on TW Hya allows us to infer constraints on the accretion geometry. The pre-shock accretion velocity is directly probed by the post-shock temperature. The EMD clearly indicates a cool peak at $T\sim2.5-3$\,MK \citep{KastnerHuenemoerder2002,BrickhouseCranmer2010}. Assuming strong-shock conditions, the predicted post-shock temperature $T_{post}$ depends only on the pre-shock velocity, i.e. $T_{post} = (3\,\mu\,m_{\rm p}\,v_{pre}^2)/(16\,k_{\rm B})$, where $\mu\,m_{\rm p}$ is the average particle mass, whose typical value is $\sim0.6\,m_{\rm p}$ \citep{ArgiroffiMaggio2007}. Even considering the complex and variable structure of the post-shock plasma, this relation holds \citep{SaccoOrlando2010}. Therefore the peak temperature of the EMD suggests an intrinsic pre-shock velocity of $430-470\,{\rm km\,s^{-1}}$. This inferred velocity fits neatly between the maximum value provided by the free-fall velocity  of $\sim500\,{\rm km\,s^{-1}}$ that characterizes TW~Hya, and the minimum, $\sim300\,{\rm km\,s^{-1}}$, needed to have an X-ray emitting post-shock plasma \citep{SaccoOrlando2010}. 

\begin{figure}
\centering
\includegraphics[width=8.5cm]{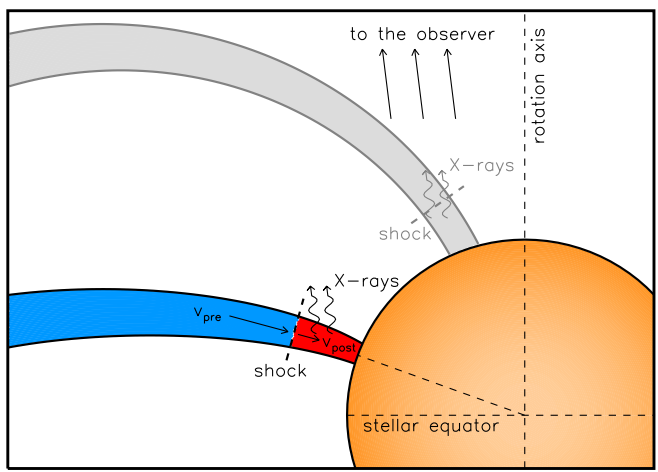}
\caption{Schematic illustration of the low-latitude accretion-shock position on the stellar surface of TW~Hya, as inferred from the observed X-ray Doppler shift. The red area at the base of the accretion stream indicates the post-shock region, where the high-density plasma at temperature of $2-4$\,MK, responsible for the emission of cool redshifted X-ray lines, is located. Accretion streams terminating near the stellar pole (as that shown in gray) cannot be excluded, though their X-ray post-shock emission could not in fact be readily observed because of the absorption within the pre-shock material.}
\label{fig:accretioncartoon}
\end{figure}

The X-ray emitting post-shock plasma is expected to show in its X-ray emission a bulk inward velocity of $v_{post}\approx v_{pre}/4 \approx 110-120\,{\rm km\,s^{-1}}$, even taking into account the complex spatial structure of the post-shock and its intrinsic instabilities (Bonito et al., in prep.). Then, because of the radial velocity of $\sim40\,{\rm km\,s^{-1}}$ that we detected, we can infer an inclination angle between the post-shock velocity direction and line of sight of $\sim70^{\circ}$. And hence, assuming that the accretion streams arrive perpendicularly with respect to the stellar surface, the base of the accretion streams producing the observed soft X-ray emission from TW~Hya appear to be located at low latitudes. Considering that the inclination of TW~Hya is low, but not zero, and assuming $i=7^{\circ}$ \citep{QiHo2004}, we can then infer that the latitude of these X-ray emitting accretion shocks should lie between $10^{\circ}$ and $30^{\circ}$, as shown in Fig.~\ref{fig:accretioncartoon}. Summarizing, the constraint provided by our analysis is stringent because we have both an inference as to the intrinsic pre-shock velocity, based on the temperature of the X-ray emitting post-shock plasma, and a measurement of the post-shock velocity along the line of sight measured by the redshift. Finally, even guessing that the observed velocity of $\sim40\,{\rm km\,s^{-1}}$ comes out from a mixture of plasma components with radial velocities ranging, for instance, from 0 to $\sim80\,{\rm km\,s^{-1}}$ (because of the spectral resolution of the {\it Chandra}/HETGS), the corresponding latitudes would be lower than $50^{\circ}$, confirming also in this case that X-ray emitting accretion spots are mainly located at low latitudes.

The existence of significant accretion terminating at low latitudes confirms that the geometry of the accretion streams depends not only on the dipole component, but on the magnetic configuration near the stellar surface \citep{GregoryJardine2006} that is characterized on TW~Hya by intense higher order magnetic components \citep{DonatiGregory2011}.

Because the start and stop times of the TW~Hya~(1) observation cover a period of $\sim$15\,d and because the redshift appears to be the same on shorter time intervals as well, these low-latitude accretion spots seem to be stable over timescales of a few weeks. Moreover, despite the lower significance level, the same radial velocity also emerges from the TW~Hya~(2) observation, indicating that the geometry of these low-latitude accretion spots could likely have remained stable over a timescale of a few years. Evidence of long-term stability of the accretion configuration was observed in a few other cases \citep[e.g.,][]{HamaguchiGrosso2012,SiciliaAguilarFang2015}, although accretion more often appears as a highly variable phenomenon \citep[e.g.,][]{SousaAlencar2016,RigonScholz2017}.

Even if the observed X-ray redshift from TW Hya indicates the existence of an apparently stable low-latitude accretion-shock area, X-ray data alone cannot exclude the existence of accretion streams terminating near the stellar pole. X-rays are in fact very easily absorbed by the pre-shock material \citep{BonitoOrlando2014,CostaOrlando2017}, hence post-shock X-rays are not expected to be easily observed when a significant portion of pre-shock material is located along the line of sight \citep{ArgiroffiFlaccomio2011}. Because of the almost pole-on view of TW~Hya, this could be the case for streams terminating at very high latitudes (Fig.~\ref{fig:accretioncartoon}).

\subsection{Comparison with other constraints on accretion geometry}

In this work we focused on the X-ray emission due to the post-shock plasma on TW~Hya, measuring a redshift of $38.3\pm5.1\,{\rm km\,s^{-1}}$ and inferring that the accretion-shock regions responsible for this emission should be located at low latitude on the stellar surface. In addition to the plasma heated in the accretion shocks, accretion in low-mass stars involves several components, from cold inner disk structures to accretion streams. To try to depict a coherent picture of the accretion process on TW~Hya, we compare here our results to those previously obtained exploiting different diagnostics and/or focusing on different accretion components.
The \ion{C}{iv} line doublets at 1550\,\AA\, probes plasma at $\sim10^{5}$\,K involved in the accretion process. In CTTS this doublet usually shows a profile characterized by a narrow component (NC) and a broad component (BC). \citet{ArdilaHerczeg2013} suggested that the NC originates in the post-shock region, where the material cools down after having been heated up to few MK, while the BC instead forms in pre-shock portions of the accretion streams. Both the NC and BC of the \ion{C}{iv} line profiles of TW~Hya are redshifted with peaks located at $\approx30$ and $\approx120\,{\rm km\,s^{-1}}$, respectively \citep{ArdilaHerczeg2013}. The peak of the NC is very similar to the velocity we measured for the plasma at 3\,MK. This strongly suggests that both \ion{C}{iv} NC and X-ray cool lines originate in the same post-shock plasma with the $10^5$\,K material located deeper in the post-shock cooling zone, as predicted by MHD modeling \citep{ColomboOrlando2016}. Therefore, the \ion{C}{iv} line doublet also points to low-latitude accretion spots on TW~Hya.

The absorption cross-section in the soft X-ray band ($\sim$15$-$20\,\AA) is similar to that in the far UV band \citep[$\sim1000-2000$\,\AA, ][]{Ryter1996}. Therefore, as for the X-rays, UV line photons emitted by the post-shock plasma can easily escape sideways with respect to the stream direction, but are highly absorbed if emitted in the stream direction. Therefore, both UV and soft X-ray lines preferentially probe streams observed in the direction perpendicular to the stream. In the case of TW~Hya, therefore, UV emission confirms low-latitude spots, but cannot exclude a polar shock region. Considering in fact the large sample of CTTS inspected by \citet{ArdilaHerczeg2013}, the \ion{C}{iv} NC shows on average low radial velocities. This could be because this bias is related to the post-shock observability that is related to the stream direction, and not because accretion inflow motions are slower than the free-fall velocity. The MK temperature of the post-shock guarantees, in fact, that the intrinsic pre-shock velocity should be as high as that predicted by free fall, hence the low radial velocity observed on average for the \ion{C}{iv} NC \citep{ArdilaHerczeg2013} indicates that the emitting accretion regions are viewed at high inclination with respect to the stream direction.

A low-latitude accretion spot on TW~Hya was previously suggested by \citet{BatalhaBatalha2002} to interpret veiling and line variations as due to rotational modulation of the hot spot. Moreover the existence of equatorial accretion streams agrees with the magnetic topology of TW~Hya, characterized by higher order components that dominate the dipolar component \citep{DonatiGregory2011}.

As discussed above, the observed X-ray redshift points to a low-latitude stream, but cannot exclude the presence of polar streams. Their existence is indeed suggested by the profile of the \ion{He}{I} line at 10380\,\AA. This line clearly shows a subcontinuum redshifted absorption feature extending up to $\sim300-400\,{\rm km\,s^{-1}}$ \citep{FischerKwan2008,DupreeBrickhouse2014}. This absorbing material can be obtained only with a free-fall velocity oriented along the line of sight, as is precisely the case for an accretion funnel terminating at high latitude. \citet{DonatiGregory2011}, performing  spectropolarimetric monitoring of the \ion{Ca}{II} infrared triplet, and interpreting the line variability as due to rotational modulation, found that both polar and equatorial accretion indeed occur on TW~Hya.

Accretion can also be probed by hydrogen Balmer lines. However, interpreting their profiles to infer constraints on the accretion geometry is nontrivial. In TW~Hya both H$\alpha$ and H$\beta$ lines present a redshifted absorption feature at $\sim35\,{\rm km\,s^{-1}}$ \citep{AlencarBatalha2002,DupreeBrickhouse2014}. This feature appears to be stable in observations performed at different epochs, and it is present also during observations performed simultaneously with the 2007 {\it Chandra} observing campaign. This absorption is usually interpreted as due to the hot-spot emission observed through the accretion column, which implies an accretion stream terminating at high latitude \citep{AlencarBatalha2002}. However, because of the pole-on inclination of TW~Hya, a polar stream would probably produce absorption features at higher velocity. Moreover, this feature provides a redshifted velocity that is similar to that observed from soft X-rays and far UV lines. For this reason, we speculate that models predicting H$\alpha$ and H$\beta$ line emission, both from within the outer stellar atmosphere and in post- and pre-shock material, are needed to test whether or not the similar redshifts shown by optical absorption features, far UV lines, and soft X-rays lines, indicate that they all originate from the same structure.

\section{Conclusions}

Thanks to the wavelength resolution and calibration of the {\it Chandra}/HETGS gratings, we measured for the first time the redshift, with respect to the stellar photosphere, of the cool X-ray emitting plasma component of the CTTS TW~Hya. The observed redshift of $38.3\pm5.1\,{\rm km\,s^{-1}}$ indicates the following:

\begin{itemize}

\item[-] This X-ray emitting plasma at 2$-$4\,MK is receding with respect to the photosphere and is indeed located in the post-shock region at the base of the accretion streams, as predicted by several HD and MHD models, strongly confirming the nature of the high-density X-ray emitting plasma component observed in general in CTTS.

\item[-] The footpoints of these X-ray emitting accretion streams are located at low latitudes ($\sim10^{\circ}-30^{\circ}$) on the stellar surface, as strictly constrained by comparing the measured redshift with the intrinsic pre-shock velocity required to generate the observed temperature of the post-shock X-ray emitting region.

\item[-] The observed velocity is very similar to that of the NC of the C~IV resonance doublet, suggesting that both soft X-rays and far UV lines are emitted by the same low-latitude post-shock regions.

\end{itemize}

This new detection of the X-ray redshift from material accreting onto a CTTS, in addition to providing constraints on the nature of the X-ray emitting plasma and on the magnetospheric accretion geometry, is also important in light of the future X-ray mission {\it Athena}, where the combination of large effective area and high spectral resolution will allow us to extend this diagnostic to a larger and more diverse population of CTTS \citep{NandraBarret2013,BarretLamTrong2016}.

\begin{acknowledgements}
We thank the referee, Joel Kastner, for very constructive comments. The scientific results reported in this article are based on observations obtained from the Chandra Data Archive.
\end{acknowledgements}

\bibliographystyle{aa} 
\bibliography{twhya_xds}

\begin{appendix}

\section{Line Doppler shifts}

\begin{landscape}

\begin{figure}
\centering
\includegraphics[width=25cm]{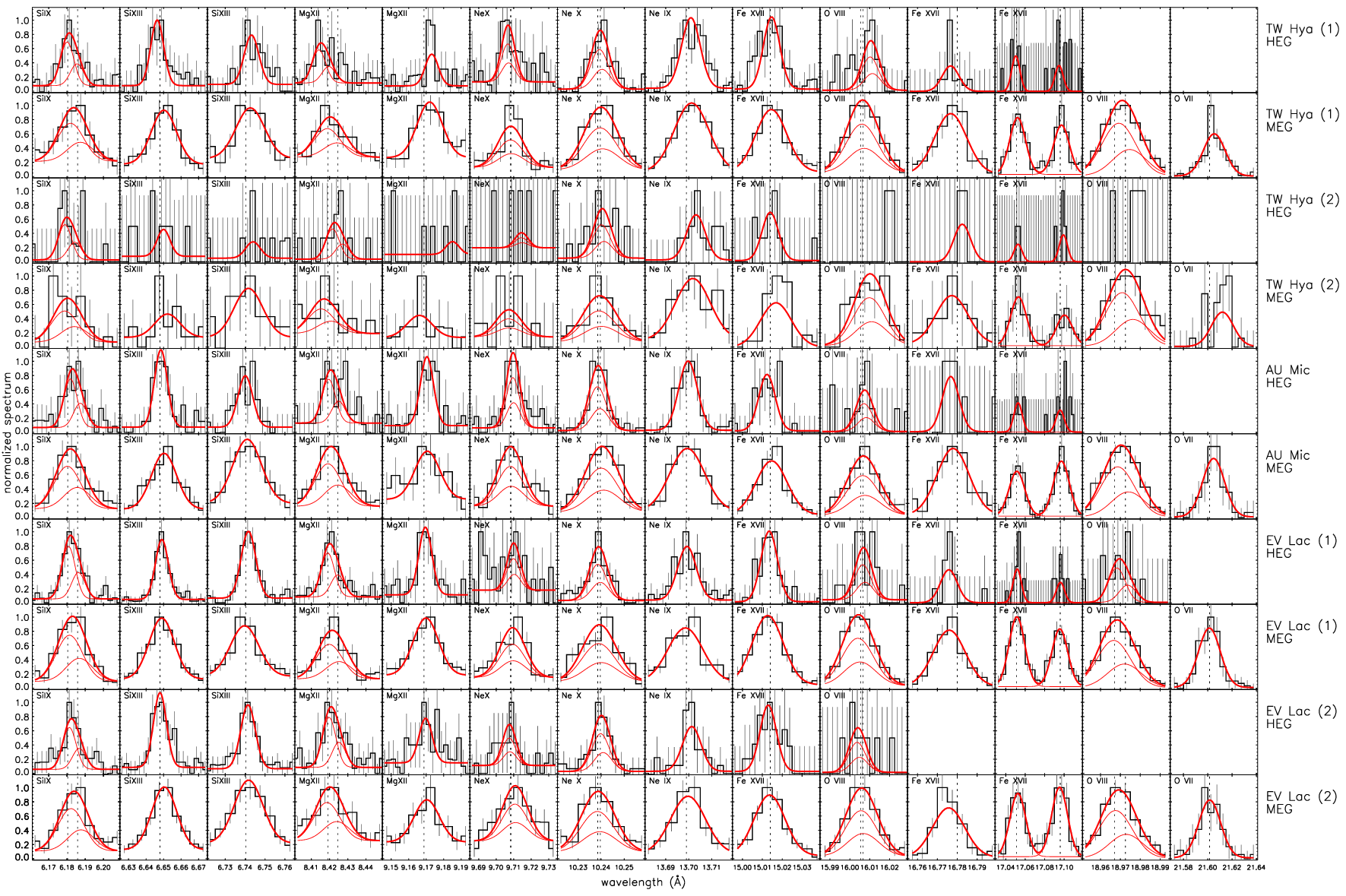}
\caption{Observed HEG and MEG spectra of TW~Hya~(1), TW~Hya~(2), AU~Mic, EV~Lac~(1), and EV~Lac~(2), in the regions of the selected emission lines (black) with the best-fit functions superimposed (thick red line). In the case of resonance doublets, we plot also the individual Gaussian components (thin red line). Dashed vertical lines indicate the positions of the predicted wavelengths.}
\label{fig:fitlines1}
\end{figure}

\end{landscape}

\begin{landscape}

\begin{figure}
\centering
\includegraphics[width=25cm]{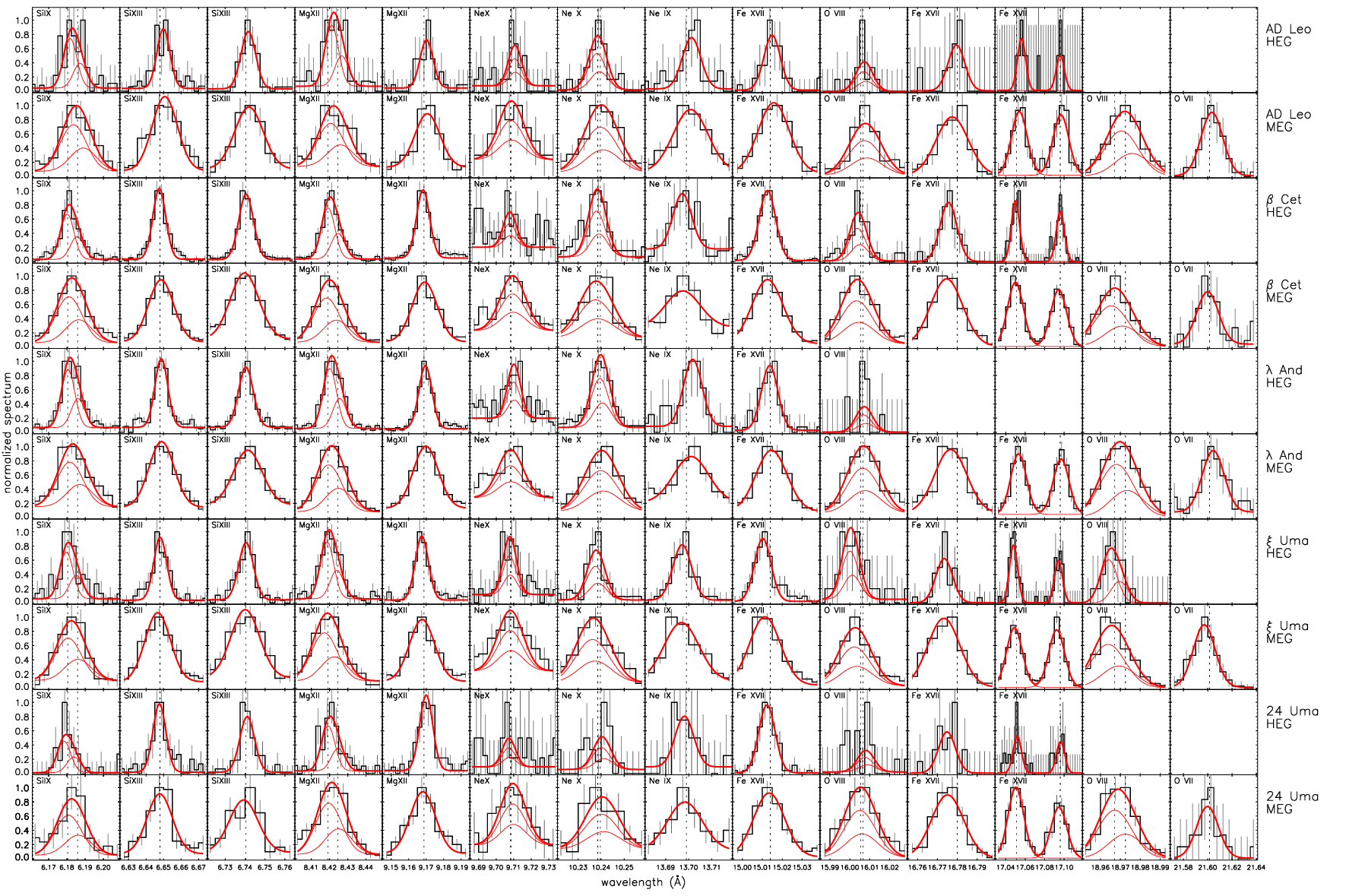}
\caption{Observed HEG and MEG spectra of AD~Leo, $\beta$~Cet, $\lambda$~And, $\xi$~Uma, and 24~Uma, in the regions of the selected emission lines (black) with the best-fit functions superimposed (thick red line). In the case of resonance doublets, we plot also the individual Gaussian components (thin red line). Dashed vertical lines indicate the positions of the predicted wavelengths.}
\label{fig:fitlines2}
\end{figure}

\end{landscape}

We show in Figs.~\ref{fig:fitlines1} and \ref{fig:fitlines2} the observed and fitted line profiles, for both HEG and MEG gratings, for all the inspected observations. The velocities corresponding to the displacement of each line with respect to its predicted rest wavelength are listed in Table~\ref{tab:linewvl}. We also report in Table~\ref{tab:linewvl} the cool and hot plasma velocities of each observation, obtained by computing the weighted average of the velocities of cool and hot line subsamples, as classified in Table~\ref{tab:linesel}.

\begin{sidewaystable*}
\caption{Line Doppler shifts.}
\label{tab:linewvl}
\small
\begin{center}
\begin{tabular}{lrcr@{$\,\pm\,$}lr@{$\,\pm\,$}lr@{$\,\pm\,$}lr@{$\,\pm\,$}lr@{$\,\pm\,$}lr@{$\,\pm\,$}lr@{$\,\pm\,$}lr@{$\,\pm\,$}lr@{$\,\pm\,$}lr@{$\,\pm\,$}l}
\hline\hline
 & & &  \multicolumn{2}{c}{TW Hya (1)} & \multicolumn{2}{c}{TW Hya (2)} & \multicolumn{2}{c}{AU Mic} & \multicolumn{2}{c}{EV Lac (1)} & \multicolumn{2}{c}{EV Lac (2)} & \multicolumn{2}{c}{AD Leo} & \multicolumn{2}{c}{$\beta$ Cet} & \multicolumn{2}{c}{$\lambda$ And} & \multicolumn{2}{c}{$\xi$ Uma} & \multicolumn{2}{c}{24 Uma} \\
 Elem. & $\lambda$\,(\AA) & grat. &  \multicolumn{2}{c}{(${\rm km\,s^{-1}}$)} & \multicolumn{2}{c}{(${\rm km\,s^{-1}}$)} & \multicolumn{2}{c}{(${\rm km\,s^{-1}}$)} & \multicolumn{2}{c}{(${\rm km\,s^{-1}}$)} & \multicolumn{2}{c}{(${\rm km\,s^{-1}}$)} & \multicolumn{2}{c}{(${\rm km\,s^{-1}}$)} & \multicolumn{2}{c}{(${\rm km\,s^{-1}}$)} & \multicolumn{2}{c}{(${\rm km\,s^{-1}}$)} & \multicolumn{2}{c}{(${\rm km\,s^{-1}}$)} & \multicolumn{2}{c}{(${\rm km\,s^{-1}}$)} \\
\hline
\multirow{2}{*}{\ion{Si}{XIV}     } & \multirow{2}{*}{  6.1804   } & HEG &$ -15$ & $  66$ & $ -76$ & $ 169$ & $  72$ & $  63$ & $  21$ & $  31$ & $  43$ & $  47$ & $  64$ & $  74$ & $ -13$ & $  26$ & $  10$ & $  24$ & $ -36$ & $  51$ & $ -99$ & $  81$ \\
                                    &                              & MEG &$  75$ & $  81$ & $ -83$ & $ 209$ & $  -9$ & $  63$ & $  46$ & $  37$ & $  88$ & $  40$ & $ 150$ & $  68$ & $  31$ & $  27$ & $  57$ & $  30$ & $  12$ & $  52$ & $  20$ & $  87$ \\
\hline
\multirow{2}{*}{\ion{Si}{XIII}    } & \multirow{2}{*}{  6.6479   } & HEG &$ -71$ & $  53$ & $  91$ & $ 406$ & $  22$ & $  41$ & $  48$ & $  24$ & $   0$ & $  34$ & $  88$ & $  48$ & $ -18$ & $  15$ & $  34$ & $  16$ & $   5$ & $  26$ & $ -15$ & $  38$ \\
                                    &                              & MEG &$ 100$ & $  54$ & $ 194$ & $ 330$ & $ 104$ & $  46$ & $  45$ & $  30$ & $ 109$ & $  30$ & $ 124$ & $  42$ & $  27$ & $  16$ & $  40$ & $  21$ & $ -57$ & $  28$ & $  -6$ & $  43$ \\
\hline
\multirow{2}{*}{\ion{Si}{XIII}    } & \multirow{2}{*}{  6.7403   } & HEG &$ 130$ & $  77$ & $ 163$ & $ 425$ & $  -8$ & $  59$ & $  60$ & $  28$ & $  51$ & $  29$ & $  58$ & $  38$ & $   8$ & $  13$ & $   3$ & $  18$ & $   7$ & $  28$ & $  32$ & $  39$ \\
                                    &                              & MEG &$ 103$ & $  75$ & $  62$ & $ 174$ & $  33$ & $  61$ & $ -25$ & $  39$ & $  65$ & $  45$ & $  72$ & $  52$ & $ -25$ & $  20$ & $  46$ & $  26$ & $ -11$ & $  38$ & $ -46$ & $  60$ \\
\hline
\multirow{2}{*}{\ion{Mg}{XII}     } & \multirow{2}{*}{  8.4192   } & HEG &$-178$ & $  68$ & $  88$ & $ 185$ & $  13$ & $  70$ & $  -5$ & $  34$ & $  20$ & $  41$ & $  71$ & $  56$ & $   1$ & $  11$ & $  34$ & $  10$ & $ -20$ & $  22$ & $   2$ & $  34$ \\
                                    &                              & MEG &$ -15$ & $  63$ & $-137$ & $ 252$ & $  -1$ & $  42$ & $  23$ & $  41$ & $ -22$ & $  46$ & $  49$ & $  50$ & $ -29$ & $  11$ & $  11$ & $   9$ & $ -72$ & $  27$ & $  13$ & $  25$ \\
\hline
\multirow{2}{*}{\ion{Mg}{XI}      } & \multirow{2}{*}{  9.1687   } & HEG &$ 145$ & $  52$ & $ 534$ & $ 598$ & $  55$ & $  48$ & $  26$ & $  35$ & $  30$ & $  55$ & $  49$ & $  46$ & $ -16$ & $  10$ & $  27$ & $  10$ & $ -39$ & $  14$ & $  41$ & $  20$ \\
                                    &                              & MEG &$ 108$ & $  48$ & $ -83$ & $ 241$ & $  63$ & $  77$ & $  37$ & $  30$ & $  52$ & $  57$ & $  64$ & $  55$ & $  12$ & $  12$ & $  35$ & $  12$ & $ -34$ & $  16$ & $ -21$ & $  26$ \\
\hline
\multirow{2}{*}{\ion{Ne}{X}       } & \multirow{2}{*}{  9.7080   } & HEG &$ -41$ & $  57$ & $ 196$ & $ 979$ & $  46$ & $  36$ & $  57$ & $  75$ & $ -18$ & $  73$ & $  78$ & $  77$ & $ -10$ & $  59$ & $  60$ & $  31$ & $ -12$ & $  44$ & $ -26$ & $ 147$ \\
                                    &                              & MEG &$   5$ & $  47$ & $ -28$ & $ 385$ & $   3$ & $  50$ & $  39$ & $  38$ & $  82$ & $  50$ & $  13$ & $  92$ & $  42$ & $  29$ & $   1$ & $  29$ & $   0$ & $  47$ & $  44$ & $  79$ \\
\hline
\multirow{2}{*}{\ion{Ne}{X}       } & \multirow{2}{*}{ 10.2385   } & HEG &$  22$ & $  24$ & $  48$ & $  89$ & $  -2$ & $  24$ & $   1$ & $  23$ & $  37$ & $  28$ & $  -9$ & $  33$ & $ -15$ & $  22$ & $  27$ & $  15$ & $ -29$ & $  23$ & $  45$ & $  89$ \\
                                    &                              & MEG &$  30$ & $  21$ & $  15$ & $  72$ & $  44$ & $  25$ & $  12$ & $  24$ & $  -3$ & $  25$ & $  35$ & $  31$ & $ -28$ & $  19$ & $  39$ & $  12$ & $ -65$ & $  20$ & $  45$ & $  57$ \\
\hline
\multirow{2}{*}{\ion{Ne}{IX}      } & \multirow{2}{*}{ 13.6990   } & HEG &$  43$ & $  17$ & $  92$ & $  51$ & $  16$ & $  22$ & $   6$ & $  23$ & $  51$ & $  36$ & $  49$ & $  39$ & $ -38$ & $  28$ & $  60$ & $  31$ & $ -39$ & $  17$ & $ -18$ & $  67$ \\
                                    &                              & MEG &$  50$ & $  11$ & $  62$ & $  44$ & $   9$ & $  17$ & $  -8$ & $  18$ & $  16$ & $  19$ & $  40$ & $  21$ & $ -34$ & $  27$ & $  45$ & $  22$ & $ -45$ & $  14$ & $ -13$ & $  36$ \\
\hline
\multirow{2}{*}{\ion{Fe}{XVII}    } & \multirow{2}{*}{ 15.0140   } & HEG &$  13$ & $  22$ & $   0$ & $  78$ & $ -32$ & $  34$ & $ -10$ & $  19$ & $ -20$ & $  43$ & $  18$ & $  26$ & $ -28$ & $   7$ & $  -6$ & $  21$ & $ -74$ & $  11$ & $ -33$ & $  12$ \\
                                    &                              & MEG &$  10$ & $  13$ & $  57$ & $  45$ & $  13$ & $  14$ & $ -25$ & $  13$ & $  -6$ & $  13$ & $  35$ & $  15$ & $ -27$ & $   5$ & $  12$ & $   9$ & $ -53$ & $   8$ & $ -11$ & $   9$ \\
\hline
\multirow{2}{*}{\ion{O}{VIII}     } & \multirow{2}{*}{ 16.0055   } & HEG &$  88$ & $  40$ & \multicolumn{2}{c}{$\cdot\cdot\cdot$} & $  30$ & $  79$ & $  18$ & $  57$ & $ -34$ & $ 112$ & $  26$ & $  61$ & $ -30$ & $  20$ & $  26$ & $  89$ & $-101$ & $  38$ & $  40$ & $  94$ \\
                                    &                              & MEG &$  10$ & $  17$ & $  80$ & $  43$ & $  18$ & $  17$ & $ -27$ & $  12$ & $   1$ & $  17$ & $  36$ & $  18$ & $ -38$ & $   9$ & $  19$ & $  11$ & $ -59$ & $  11$ & $  -9$ & $  19$ \\
\hline
\multirow{2}{*}{\ion{Fe}{XVII}    } & \multirow{2}{*}{ 16.7800   } & HEG &$ -59$ & $  65$ & $  45$ & $ 221$ & $ -58$ & $  83$ & $ -74$ & $  56$ & \multicolumn{2}{c}{$\cdot\cdot\cdot$} & $  -7$ & $  71$ & $ -78$ & $  19$ & \multicolumn{2}{c}{$\cdot\cdot\cdot$} & $-116$ & $  25$ & $ -91$ & $  47$ \\
                                    &                              & MEG &$ -58$ & $  19$ & $ -47$ & $  91$ & $ -36$ & $  24$ & $ -71$ & $  15$ & $ -77$ & $  18$ & $ -44$ & $  21$ & $ -95$ & $   6$ & $ -58$ & $  13$ & $-119$ & $   8$ & $ -87$ & $  13$ \\
\hline
\multirow{2}{*}{\ion{Fe}{XVII}    } & \multirow{2}{*}{ 17.0510   } & HEG &$ -11$ & $  98$ & $  29$ & $ 234$ & $  29$ & $  87$ & $  14$ & $  50$ & \multicolumn{2}{c}{$\cdot\cdot\cdot$} & $  92$ & $  91$ & $ -17$ & $  16$ & \multicolumn{2}{c}{$\cdot\cdot\cdot$} & $ -52$ & $  21$ & $  15$ & $  43$ \\
                                    &                              & MEG &$  20$ & $  18$ & $  37$ & $  47$ & $   7$ & $  19$ & $   8$ & $  12$ & $  26$ & $  21$ & $  51$ & $  20$ & $ -18$ & $   5$ & $  39$ & $  11$ & $ -36$ & $   8$ & $  -9$ & $  10$ \\
\hline
\multirow{2}{*}{\ion{Fe}{XVII}    } & \multirow{2}{*}{ 17.0960   } & HEG &$ -20$ & $ 125$ & $  57$ & $ 155$ & $   0$ & $ 110$ & $   4$ & $  80$ & \multicolumn{2}{c}{$\cdot\cdot\cdot$} & $   1$ & $ 123$ & $   2$ & $  17$ & \multicolumn{2}{c}{$\cdot\cdot\cdot$} & $ -32$ & $  26$ & $   6$ & $  49$ \\
                                    &                              & MEG &$  23$ & $  19$ & $  76$ & $  72$ & $  21$ & $  17$ & $  -6$ & $  14$ & $ -10$ & $  20$ & $  12$ & $  19$ & $ -27$ & $   6$ & $  15$ & $  11$ & $ -63$ & $   7$ & $ -23$ & $  12$ \\
\hline
\multirow{2}{*}{\ion{O}{VIII}     } & \multirow{2}{*}{ 18.9671   } & HEG &\multicolumn{2}{c}{$\cdot\cdot\cdot$} & \multicolumn{2}{c}{$\cdot\cdot\cdot$} & \multicolumn{2}{c}{$\cdot\cdot\cdot$} & $  18$ & $  76$ & \multicolumn{2}{c}{$\cdot\cdot\cdot$} & \multicolumn{2}{c}{$\cdot\cdot\cdot$} & \multicolumn{2}{c}{$\cdot\cdot\cdot$} & \multicolumn{2}{c}{$\cdot\cdot\cdot$} & $ -46$ & $  38$ & \multicolumn{2}{c}{$\cdot\cdot\cdot$} \\
                                    &                              & MEG &$  33$ & $   7$ & $  60$ & $  22$ & $  24$ & $   9$ & $  -6$ & $   7$ & $   7$ & $   8$ & $  54$ & $   9$ & $ -25$ & $   8$ & $  15$ & $   6$ & $ -50$ & $   6$ & $   1$ & $  13$ \\
\hline
\multirow{2}{*}{\ion{O}{VII}      } & \multirow{2}{*}{ 21.6015   } & HEG &\multicolumn{2}{c}{$\cdot\cdot\cdot$} & \multicolumn{2}{c}{$\cdot\cdot\cdot$} & \multicolumn{2}{c}{$\cdot\cdot\cdot$} & \multicolumn{2}{c}{$\cdot\cdot\cdot$} & \multicolumn{2}{c}{$\cdot\cdot\cdot$} & \multicolumn{2}{c}{$\cdot\cdot\cdot$} & \multicolumn{2}{c}{$\cdot\cdot\cdot$} & \multicolumn{2}{c}{$\cdot\cdot\cdot$} & \multicolumn{2}{c}{$\cdot\cdot\cdot$} & \multicolumn{2}{c}{$\cdot\cdot\cdot$} \\
                                    &                              & MEG &$  53$ & $  16$ & $ 146$ & $  59$ & $  45$ & $  28$ & $  -2$ & $  17$ & $   1$ & $  26$ & $  26$ & $  25$ & $ -22$ & $  37$ & $  41$ & $  27$ & $ -61$ & $  16$ & $ -22$ & $  60$ \\
\hline
\hline
\multicolumn{          23}{c}{Average velocities of the X-ray emitting cool and hot plasma components}\\
\hline
\multicolumn{3}{c}{$v_{\rm Xc}$} & $  38.9$ & $   5.1$ & $  73.1$ & $  16.4$ & $  21.5$ & $   6.5$ & $  -9.9$ & $   5.3$ & $   8.2$ & $   6.6$ & $  46.6$ & $   7.2$ & $ -31.5$ & $   5.4$ & $  18.9$ & $   4.7$ & $ -52.7$ & $   4.6$ & $  -3.9$ & $   9.9$ \\
\multicolumn{3}{c}{$v_{\rm Xh}$} & $  19.9$ & $   6.9$ & $  38.8$ & $  23.2$ & $  17.4$ & $   7.1$ & $   5.9$ & $   5.3$ & $  16.9$ & $   6.9$ & $  40.1$ & $   7.6$ & $ -19.1$ & $   2.2$ & $  24.8$ & $   3.2$ & $ -48.9$ & $   3.4$ & $ -11.4$ & $   4.7$ \\
\hline
\end{tabular}
\end{center}
All the velocities are in the {\it Chandra} reference frame. All the errors are at 1\,$\sigma$ level.
\normalsize
\end{sidewaystable*}

\end{appendix}

\end{document}